\documentstyle[12pt,preprint]{aastex}
\newcommand\be{\begin{equation}}
\newcommand\ee{\end{equation}}
\newcommand\eq{\begin{equation}}
\newcommand\en{\end{equation}}
\newcommand\bx{{\bf x}}
\newcommand\bu{{\bf u}}

\begin{document}

\title{The Effects of Massive Substructures on Image Multiplicities in 
Gravitational Lenses}

\author{ J.D. Cohn${}^1$ and C.S. Kochanek${}^2$}
\affil{${}^1$ Space Sciences Laboratory and
Theoretical Astrophysics Center,}
\affil{ 601 Campbell Hall, UC Berkeley, Berkeley, CA 94720-3411, USA}
\affil{${}^2$Harvard-Smithsonian Center for Astrophysics, 60 Garden St.,
  Cambridge, MA 02138}
\affil{email: jcohn@astron.berkeley.edu, ckochanek@cfa.harvard.edu}

\begin{abstract}
Surveys for gravitational lens systems have typically found a significantly larger 
fraction of lenses with four (or more) images than are predicted by standard 
ellipsoidal lens models (50\% versus 25-30\%).  We show that including the
effects of smaller satellite galaxies, with an abundance normalized by the
observations, significantly increases the expected number of systems with 
more than
two images and largely explains the discrepancy.  The effect is dominated by
satellites with $\sim 20$\% the luminosity of the primary lens, in rough
agreement with the typical luminosities of the observed satellites.  We
find that the lens systems with satellites cannot  be dropped from
estimates of the cosmological model based on gravitational lens statistics
without significantly biasing the results.  
\end{abstract}

\section{Introduction}

While many gravitational lenses can be modeled as single, isolated,
massive galaxies, this can only be an approximation.  Both the
luminosity functions of galaxies and the mass functions of halos
derived from hierarchical structure formation
predict that massive galaxies are likely to be surrounded by
lower mass halos.  These halos modify the gravitational potential from
that of the central (usually) elliptical galaxy and
have several observational consequences.
First, the satellite galaxies observed in many systems (e.g. MG0414+0534,
Schechter \& Moore \cite{SchMoo93}; B1030+074, Xanthopoulos et al. 
\cite{xanetal98};
B1152+199, Myers et al. \cite{Mye99}, 
Rusin et al. \cite{Rus02}; and B1359+154, Myers et al. \cite{Mye99} 
for the JVAS/CLASS sample) 
are required to obtain a successful model of the
observed image geometries.  The gravitational field produced by
an offset nearby satellite is essentially impossible to mimic through
variations in the structure of the central lens galaxy or the
addition of tidal shears. 
Second, the satellite galaxies can
change the caustic structure of the lens, possibly helping to
explain the relatively high numbers of observed quad lenses
(e.g. Kochanek \& Apostolakis \cite{KocApo88}, 
Rusin \& Tegmark, \cite{RusTeg01}, Moller \& Blain \cite{MolBla01}).
Third, even very low mass satellites can perturb the
magnification tensors of individual images so as to produce
patterns of image magnifications which cannot be reproduced by
the central lens galaxy (Mao \& Schneider \cite{MaoSch98}, Metcalf \& Madau
\cite{MetMad01}, Chiba \cite{Chi02}, Dalal \& Kochanek \cite{DalKoc01}, 
Schechter \& Wambsganss \cite{SchWam02}, Keeton \cite{Kee03},
Keeton, Gaudi, Petters \cite{KeeGauPet02}, Kochanek \& Dalal 
\cite{KocDal02,KocDal03}, Evans \& Witt~\cite{Evans03}).

In this paper we focus on the second of these effects,  the
role of substructure in changing
the expected numbers of images, in particular the ratio of
lenses with 2 or 4 images.   Standard statistical 
models consisting of an isolated elliptical galaxy
in an external tidal field have difficulty reproducing the
observed ratio of quad to double lenses.  In particular, 50\%
(10 of 20: 10 doubles, 9 quads and 1 with six images)
of the published lenses found in the JVAS/CLASS 
surveys (Patnaik et al. \cite{Pat92},
Browne et al. \cite{Bro98,Bro02}, Wilkinson et 
al \cite{Wil98}, King et al. \cite{Kin99}, 
Myers et al. \cite{Mye95,Mye99,Mye02})  for lensed flat-spectrum
radio sources have more than two images, while the most recent models predict 
only 24--31\% (Rusin \& Tegmark, \cite{RusTeg01}, hereafter RT).
The significance of this ``quads to doubles ratio'' problem has
risen and fallen as the lens sample has grown (see King \& Browne 
\cite{KinBro96}, Kochanek \cite{Koc96b}, Keeton, Kochanek \&
Seljak \cite{kks}, Finch et al. \cite{Fin02}), but the ratio has always 
seemed uncomfortably high.  Most solutions, other than invocations
of bad luck, such as the effects of core radii, high dark matter
ellipticities or local tidal fields, are not very successful at
changing the ratio without becoming either physically implausible
or predicting large numbers of other unobserved image configurations.  

Compound lenses have been addressed
previously within different approximations.
Kochanek \& Apostolakis (\cite{KocApo88}) examined the
caustic/multiplicity and
magnification structure for two lenses of the same mass at differing
separations and redshifts.  They focused on spherical lenses but indicated
modifications which would be introduced by ellipticity as well.
Seitz \& Schneider (\cite{SeiSch92}) considered multiplicities
and magnifications for compound lenses.
Moller \& Blain (\cite{MolBla01}) calculated probabilities
and some characteristic properties for lenses at two different
redshifts but they did not consider quantitatively the role of the correlation
function in enhancing the numbers of satellites clustered with the
primary lens.  RT made a limited study of the effects of adding faint 
SIS (singular
isothermal sphere) satellites, concluding that low mass neighbors 
would have little effect. 
Here we extend their work to include higher mass neighbors as well
(in a sequence of lens mass ratios),
the correlation function of galaxies and the 
number density of satellites.   
In \S2 we outline our calculation.  In 
\S3 we discuss the results and how they change when we vary some of our 
(standard) assumptions.
In \S4 we summarize the conclusions and outline the issues needing
further study.

\section{Methodology}

We will study the statistical consequences of including a satellite
galaxy on the lensing properties of a more massive primary lens.
In order to separate the effects of the clustering from the lensing
properties of the two halos in isolation, we focus on estimates of
the ``excess'' lensing cross sections and probabilities created by
the clustering.  In \S2.1 we outline our
method for calculating the cross sections, and mathematically 
define the excess lensing cross sections and probabilities.  
In \S2.2 we determine the normalization of the density of 
satellites needed to match the observed numbers of satellites
in the JVAS/CLASS sample.  Finally, in \S2.3 we define the lens
models and methods used in our calculation.

\subsection{Excess Lensing Cross Sections and Probabilities }

We label the primary lens and the satellite by their critical radii in
isolation,
$b_0$ and $b_1$.  The satellite is always taken to be
less massive than the primary, $b_1 \leq b_0$.  We will use 
singular isothermal spheres (SIS) for our mass distribution, 
so the critical radii are related to the velocity dispersion $\sigma$
of the lens galaxy by $b=4\pi (\sigma/c)^2 (D_{LS}/D_{OS})$. 
The quantities $D_{OL}$, $D_{LS}$ and $D_{OS}$ are comoving
distances between the Observer, Lens and Source
redshifts.   For simplicity we place the lens at the median
redshift of the observed CLASS lenses, $z_l=0.63$, and the
source, $z_s=1.6$, at twice the distance corresponding to the 
lens redshift in an $\Omega_0=0.3$ flat cosmological model.
For an $L_*$ galaxy with $\sigma_*=220$~km/s
this means that   $b_0 = 0\farcs7$.  This corresponds to
a comoving separation of $5.5 h^{-1}$~kpc at the lens redshift.
For the lens model we adopt, the 
redshifts have little consequence for the results.

The distribution of satellite galaxies around the primary lens has
two parts.  There are satellites correlated (near) the primary
lens and uncorrelated satellites projected along the line of 
sight.  We assume that the luminosity or velocity dispersion
distribution of the satellites is the same for both correlated
and uncorrelated satellites.  For ease of calculation we will
project the uncorrelated satellites into an effective distribution
at the redshift of the primary lens.  In doing so we will neglect
the lensing of the background galaxy by the foreground galaxy, 
but this should be relatively unimportant given the dominant
role of the correlated galaxies at small separations.  
The comoving density of 
satellites of luminosity $L$ is described by
\begin{equation}
\label{schech}
  { dn \over dV d L } = \left(1+\xi(r) \right) 
  { n_* \over L_* }\left(\frac{L}{L_*}\right)^{\alpha}e^{-L/L_*}.  
\end{equation}  
The expression has two parts. The first part,
$\left(1+\xi(r) \right)$, describes the density of galaxies 
as a function of the comoving distance $r$ from the primary
lens.  We model the three-dimensional correlation function by
$\xi(r)=(r/r_0)^{-\gamma}$ with $\gamma\simeq 1.8$
and a comoving correlation scale of $r_0\simeq 5h^{-1}$~Mpc
(e.g. Peacock \cite{Pea01}).  The second part is a 
Schechter \cite{Sch76} luminosity function for the 
galaxies characterized by comoving density $n_*$, 
an exponent which we will assume to be $\alpha =-1$ and
a luminosity scale $L_*$.   In order to convert from 
luminosity to deflection $b$, we use the Faber-Jackson \cite{FabJac76} 
relation $L/L_*  = (\sigma/\sigma_*)^4$ with $\sigma_*=220$~km/s
to convert from luminosity to velocity dispersion.
As some of the satellite galaxies might be later type galaxies with smaller
mass-to-light ratios than the primary lens, the use of the Faber-Jackson
relation may overestimate the masses of satellites.  We 
want an expression in terms of the effective deflection
at the redshift of the primary lens, and this must take into
account the redshift dependence of the deflection.  In particular,
a galaxy producing deflection $b_1$ at the redshift of the primary
lens produces a deflection $b_2 = b_1 D_{2S}/D_{1S}$ if shifted to
redshift $z_2$ --  a perturber of the same velocity dispersion but
lower (higher) redshift produces a larger (smaller) deflection. 
After making the change of variables we find that
\eq
  {dn \over dV db_1} =  \left(1+\xi(r) \right){2 n_* \over b_1 } 
     e^{-(b_1^2/b_*^2)(D_{1S}/D_{2S})^2}
\en
where $b_1$ is the deflection at the redshift of the primary lens,
$r$ is the comoving distance from the primary lens, 
$b_*=4\pi(\sigma_*/c)^2D_{1S}/D_{OS}$ is the deflection produced
by an $L_*$ galaxy at the redshift of the primary lens, and
the factor of $D_{1S}/D_{2S}$ provides the shift in the effective
deflection scale when the redshift of the satellite differs
significantly from the primary.  In \S2.2 we will derive $n_*$
by fitting the observed frequency of satellites in the JVAS/CLASS 
lens sample.  We will consider an alternate model, based on the mass 
function of halos in cold dark matter (CDM) simulations in \S3.

The next step is to convert from a comoving volume density to a 
projected surface density per unit solid angle $d\Omega$ by integrating 
along the line of sight, where $dV = D_{O2}^2 dD_{O2} d\Omega$ and 
$D_{O2}$ is the comoving distance to the redshift of the satellite.
The correlated term is simple because the correlation length is
small enough to ignore the changes in the deflection scale, 
i.e. $D_{1S}/D_{2S} \sim 1$.  The
three-dimensional correlation function is replaced by its projection
\eq
\label{corrfunc}
        \xi_2(R_c) = \int_{-\infty}^{\infty} dz \xi(r)   =
    3.7 r_0 (R_c/r_0)^{-0.8}
\en
where $R_c$ is the projected comoving distance from the primary lens.  
It is related to the proper distance $R$ and the angular distance 
$\theta$ by $R = \theta D_{O1}^{ang} = R_c/(1+z_1)$ where
$D_{01}^{ang}=D_{01}/(1+z_1)$ is the angular diameter distance. 
Thus, the contribution from the correlated galaxies is
\eq
  \left( {dn \over d\Omega db_1}\right)_{cor} =  
    D_{01}^2 \xi_2(R_c) {2 n_* \over b_1 } 
     e^{-(b_1^2/b_*^2)}.
\en
The integral over the uncorrelated satellites is not trivially
analytic because of the distance factors in the exponential,
\eq
  \left( {dn \over d\Omega db_1}\right)_{uncorr} = { 2 n_* \over 3 b_1}
    D_{OS}^3 F(b_1/b_*) e^{-(b_1^2/b_*^2)}
\en
where $F(u)=3\exp(u^2) \int_0^1 x^2 dx \exp(-u^2/4(1-x)^2)$, $F(0)=1$ 
and $F(1) = 0.29$.  We derived the integral using the fact that 
$D_{1S}/D_{2S}=1/2(1-x)$
for $x=D_{02}/D_{OS}$ in a flat universe with the primary lens midway
between the observer and the source.  To 4\% accuracy we can approximate
the integral by  $F(u) \sim 0.978 - 2.033 u + 2.245 u^2  - 0.901 u^3$
for $0<u<1$.  The ratio of the uncorrelated to the
correlated surface density is
\eq
   {uncorr \over corr} \simeq 0.7 {D_{01} F(u)\over r_0 }
         \left( { R_c \over r_0 } \right)^{0.8},
\en  
so the correlated density dominates the surface density on projected
scales $R_c/b_{0,c} \leq F(u)^{-5/4}\sim 5$ for $L_*$ galaxies,  
where $b_{0,c}$ is the comoving length scale 
corresponding to the deflection $b_0$ of the primary lens. 
We combine these to get the total projected density of satellites
per unit comoving area $dA_c$ (or solid angle $d\Omega$)
\eq
\begin{array}{ll}
 {dn \over dA_c db_1} &= { 1 \over D_{O1}^2 } {dn \over d\Omega db_1} =
   {2 n_* \over b_1 } 
    \left[ \xi_2(R_c) + 8 {D_{01} \over 3} F(b_1/b_*) \right] 
   e^{-(b_1^2/b_*^2)} \\
&\equiv  \left[ \xi_2(R_c) + 8 {D_{01} \over 3} F(b_1/b_*) \right] 
\frac{dn_{Sch}}{db}
\end{array}
\en
Integrating over the distributions, the 
fraction of $L_*$ primary lenses having a correlated satellite
with $b_*/4 < b_1 < b_*$ within a projected comoving radius of $R_c$ is
\begin{equation}
   f_{sat} \simeq 0.028 \left( { n_* \over 0.01 h^3 \hbox{Mpc}^{-3} } \right)
       \left( { R_c   \over 10 h^{-1}\hbox{kpc} } \right)^{1.2}
       \left( { r_0 \over  5 h^{-1}\hbox{Mpc} } \right)^{1.8},
\end{equation}
while the fraction having an uncorrelated satellite is
\begin{equation}
   f_{sat} \simeq 0.012 \left( { n_* \over 0.01 h^3 \hbox{Mpc}^{-3} } \right)
       \left( { R_c   \over 10 h^{-1}\hbox{kpc} } \right)^2.
\end{equation}
The correlated satellites dominate on scales $R_c < 30h^{-1}$~kpc.

An isolated lens with critical radius $b$ has a cross section for
producing $n$ visible images of $\sigma_n(b)$ and a total cross section
$\sigma_{tot}(b)$ from the sum of these cross sections. For a
spherical lens, only $\sigma_2(b)$ is non-zero.  In theory, our
pair of lenses can produce multiple image systems with $m=3$, $5$ or
$7$ images.  However, either 1 or 2 of the images are trapped in
the cores of the lenses, strongly demagnified and hence invisible
to an observer.  For example, in the $m=7$ image systems, two images are
always trapped in the cores to leave only $n=5$ visible images.  If
we characterize the configurations by the total number of images
$m$ and the number of visible images $n$ (those not trapped in a core),
then lenses are produced with $m/n=3/2$, $3/3$, $5/3$, 
$5/4$ and $7/5$.  
We will keep track only of the numbers of visible
images $n$.  For simplicity we did not separate the 3/3 and 3/2 systems.
Some 3/3 systems were due to our use of cores; the remaining 
number were a negligible fraction of the 3/2 systems.
Thus, a pair of lenses with critical radii of $b_0$ and $b_1 \leq b_0$
separated by projected distance $R$ have a cross section for
producing $n$ visible images of $\sigma_n(b_0,b_1,R)$ and a total
cross section of $\sigma_{tot}(b_0,b_1,R)$.  

The probability
of observing a lens must include the effects of magnification
bias  (e.g. Turner, Ostriker, Gott \cite{TurOstGot84}), as 
magnification due to lensing brings more objects into
a survey sample.  If the sources have a flux distribution
$dn/dF$, then the magnification bias factor is
\eq
  B(F) =
  \left[\frac{dn}{dF}(F)\right]^{-1}
  \int \frac{dP}{dM} \frac{dM}{M} \frac{dn}{dF}\left(\frac{F}{M}\right)
\en
given the probability distribution $dP/dM$ of the image magnifications.
We assume a flux distribution of
\eq
\frac{dn}{dF}(F) \propto F^{-(\alpha +1)} \; .
\en
where $\alpha = 1.1$ for CLASS, as described in RT.
So, associated with each cross section (e.g.~$\sigma_n(b_0,b_1,R)$) is a
magnification bias factor ($B_n(b_0,b_1,R)$) and the probability
of observing a lens is the product of the two factors,
$P_n(b_0,b_1,R)=\sigma_n(b_0,b_1,R)B_n(b_0,b_1,R)$.
For a single isolated lens we have $P_n(b) = \sigma_n(b) B_n(b)$.

We are interested in the ``excess'' lensing cross section or probability
created by having lenses which are not isolated. We define the 
excess cross section associated with a satellite $b_1$ at impact parameter
$R$,
\eq
  \label{excess1}
  \sigma_{e,n}(b_0,b_1,R) = 
    \sigma_n(b_0,b_1,R)-\sigma_n(b_0)-\sigma_n(b_1) 
\en
as the difference between the cross section produced by the combined lenses 
and that produced by the same lenses in isolation.
Similarly we define the excess lensing probability
\eq
  \label{excess2}
  P_{e,n}(b_0,b_1,R) = P_n(b_0,b_1,R)-P_n(b_0)-P_n(b_1) 
\en
as the difference between the lensing probability of the combined lenses and
the isolated lenses.  
By summing over the different image multiplicities $n$ we obtain the
total excess cross section $\sigma_{e,tot}$ and probability $P_{e,tot}$.
These quantities can be computed for any lens model and for any satellite
distribution, although our labeling scheme is tied to our subsequent use of
SIS lens models.  To estimate the significance of the results, we need
only examine the fractional changes in the optical depth or probability
relative to the more massive lens, $\sigma_{e,n}(b_0,b_1,R)/\sigma_n(b_0)$
and $P_{e,n}(b_0,b_1,R)/P_n(b_0)$.

\subsection{Normalizing the Satellite Model}

The key factor in determining the consequences of satellite galaxies is 
their absolute number, which we parameterize through the value of 
$n_*$.
 For normal galaxy luminosity functions, the comoving density
is $n_* \simeq 0.01 h^3$~Mpc$^{-3}$ (e.g. Loveday \cite{Lov00}, Kochanek et al
\cite{Kocetal01}), although the exact
value for use in calculation depends on galaxy type definitions.
For our present purposes, we adopted an empirical approach of estimating
$n_*$ from the observed numbers of lenses with satellite galaxies.  This
approach has the advantage of avoiding questions about galaxy types,
    the dependence of the correlation length on galaxy types or mass and the
    extrapolation of the correlation function from large (Mpc) to small (kpc)
    scales.
The simplest approximation is that the number of observed systems
has little dependence on the presence of satellites.  We outline this
calculation, and then describe how we included the weak dependence
(see \S3) we did find.

For each lens we search a region $R < R_{lim}$ around the lens for
satellites with critical radii between that of the primary lens $b_0$
and a limiting critical radius (i.e. deflection) $b_{lim}$.
These detection thresholds in turn correspond to a luminosity range
$L_0=(b_0/b_*)^2 > L_{sat} > L_{lim}=(b_{lim}/b_*)^2$.  Given our
model for the luminosity function, the expected number of satellites
in system $i$ is $n_* E_i$, where
\eq
\label{eival}
   n_* E_i = 2\pi \int_0^{R_{lim}} R_c dR_c \int_{b_{lim}}^{b_0}
    db_1 { dn \over dA_c db_1 }
\en
is determined by the geometry and the luminosity ratios. 
In practice the integral over $b$ will also include a
contribution from the fact that increasing $b$ increases the total
number of lenses.  We correct for this 
by adding a cross section-weighting, replacing $dn/db \rightarrow
dn/db ( 1 + .31 \frac{n_*}{0.016} b + 0.47 \frac{n_*}{0.016} b^2)$ 
in Eqn.~(\ref{eival}) above.

We estimate $n_*$ by maximizing the Poisson likelihood for the
observed satellites.  For $N_{lens}$ systems in which system $i$ has
$N_i$ observed satellites inside the detection limits set by
$R_{lim}$ and $b_{lim}$, the likelihood of the observations is
\begin{equation}
   \log L = \sum_{i=1}^{N_{lens}} \left( - n_* E_i + N_i \ln n_* \right)
\end{equation}
plus constant terms.  If the total number of observed satellites is
$N_{sat}$, the maximum likelihood estimate for the density is
\begin{equation}
    n_* =  N_{sat} \left[ \sum_{i=1}^{N_{lens}} E_i \right]^{-1}
\end{equation}
with statistical uncertainties $\delta n_*/n_* = 1/\sqrt{2 N_{sat}}$.
With the correction for the increased cross section
mentioned above, the calculations become slightly
non-linear but are easily solved by iteration. 

We performed the calculation for the combined JVAS/CLASS sample of
20 lenses available in the literature.  We dropped 5 systems:
three for which HST imaging data is 
absent (B0128+437, B0445+123 and B0850+054);
one where the primary lens has not been identified (B1555+375), and
one in which the lens geometry is not understood (B2114+022).
We searched each lens out to a radius $R_{lim}$ of 2\farcs0.
Of the remaining 15 systems, 5 have satellites within 2\farcs0
of the primary lens (MG0414+0534, B1030+071, B1152+200, B1359+154, and
B1608+434) which are visible in the HST images and necessary
components for a successful lens model.  There are six satellites
in total because B1359+154 has two satellite galaxies inside its
Einstein ring.  A detailed observational analysis of the satellite 
selection function is beyond the scope of our present
analysis, so we simply explored a plausible range of selection
thresholds.  For selection thresholds of $b_1=0\farcs05$ (6 satellites),
$0\farcs10$ (5 satellites) and $0\farcs20$ (5 satellites),
we found that $n_*=(0.016\pm0.006)h^3$~Mpc$^{-3}$, $(0.019\pm0.009)h^3$~Mpc$^{-3}$
and $(0.032\pm0.015)h^3$~Mpc$^{-3}$ respectively.  We will adopt
the most conservative estimate, $n_*=0.016 h^3$~Mpc$^{-3}$, as our fiducial 
value.  For small changes in $n_*$, the results of \S3 for the expected
numbers of lenses can be scaled linearly with $n_*$.

\subsection{Lens Models and Calculations}

We model the lenses as singular isothermal spheres (SIS) characterized
by a critical radius $b$ and a core radius $s$, 
$\rho \propto b (r^2 + s^2)^{-2} $.  The core radius $s=0.01 b$ is
introduced only to avoid numerical singularities. 
The comoving scale of this core is thus always less than 55 $h^{-1}$ pc; 
in addition, the effects of these core radii on
the cross section and bias tend to cancel out while the additional
changes which depend on the lens separations are suppressed by our
use of the lens sample itself to normalize the model.  Thus these
models very closely approximate SIS models, and for simplicity we will
refer to them as SIS models.

For a 
two-dimensional lensing potential $\phi$ such that 
$\nabla^2 \phi = 2 \kappa(r)$, the surface density of the
model in units of the critical density is
\eq
\kappa(r) = \frac{b}{2}\frac{1}{\sqrt{s^2 + r^2}}
\en
We work in units where the critical radius of the primary lens 
$b_0\equiv 1$.  We consider
satellites with critical radii, relative to $b_0$, of 
$b_1/b_0=0.3$, $0.4$, $0.5$, $0.6$, $0.7$, $0.8$, $0.9$ and 
$1.0$.  Since the outer grid scales must be held fixed, we eventually
lack the resolution needed to resolve the cores of the low mass
satellites ($b_1 \le 0.3$), forcing us extrapolate the cross sections
to smaller masses.  Since the masses roughly scale as $b^2$, we 
probe mass ratios from 1:1 to roughly 1:11.  

For each image configuration we solve the lens equations
\eq
\label{lenseq}
 \bu = \bx - \nabla \phi (\bx)
\en
to determine the image positions $\bx_i$ corresponding to a grid of source 
positions $\bu$.  The image magnification $|M|$ is analytically derived 
from the inverse magnification tensor,
\eq
 M^{-1} = I - 
         \left(
            \begin{array}{rr}
                \phi,_{xx} &\phi,_{xy} \\
                \phi,_{xy} &\phi,_{yy}
            \end{array}
         \right).
\en
Since we generally dealt with two close potentials, we neglected any external
(tidal) shears from other sources.  While such perturbations are expected
in all lenses (e.g. Kovner \cite{kovner}, Bar-Kana  \cite{barkana}), 
and seem to be
necessary components of any realistic lens model (Keeton, Kochanek \& Seljak
\cite{kks}),  the perturbations due to the two satellites are generally large
enough to neglect those from more distant objects.  However, neglecting
the tidal fields from larger scales does force us to truncate some integrals
as we discuss in detail later on.

We perform the calculation using a variant of the triangle tessellation 
method 
(e.g. Blandford \& Kochanek \cite{BlaKoc87}) on a $9000^2$ image plane grid 
and a $1700^2$ source plane grid covering the multiple image regions of
the lens and source planes.  The image plane and source
plane grid spacings were $0.005 b_0$.
The cross sections are the number of points in the source plane which
have a given image multiplicity and the magnifications 
are the projected area of the
image plane into the source plane for triangles enclosing source plane
points of interest.  Given our grid resolution, our magnification statistics
become poor for magnifications larger than 100.  We simply truncated
the magnification probability distributions at this limit,  after testing to
see that using a higher limit would have little quantitative impact on the
results. 
These cross sections and magnifications were combined with the luminosity
functions for the CLASS survey, an estimate of the number of satellites
as a function of mass, and the correlation function with the satellites
in order to calculate the change in lensing cross sections.

\begin{figure}
\plottwo{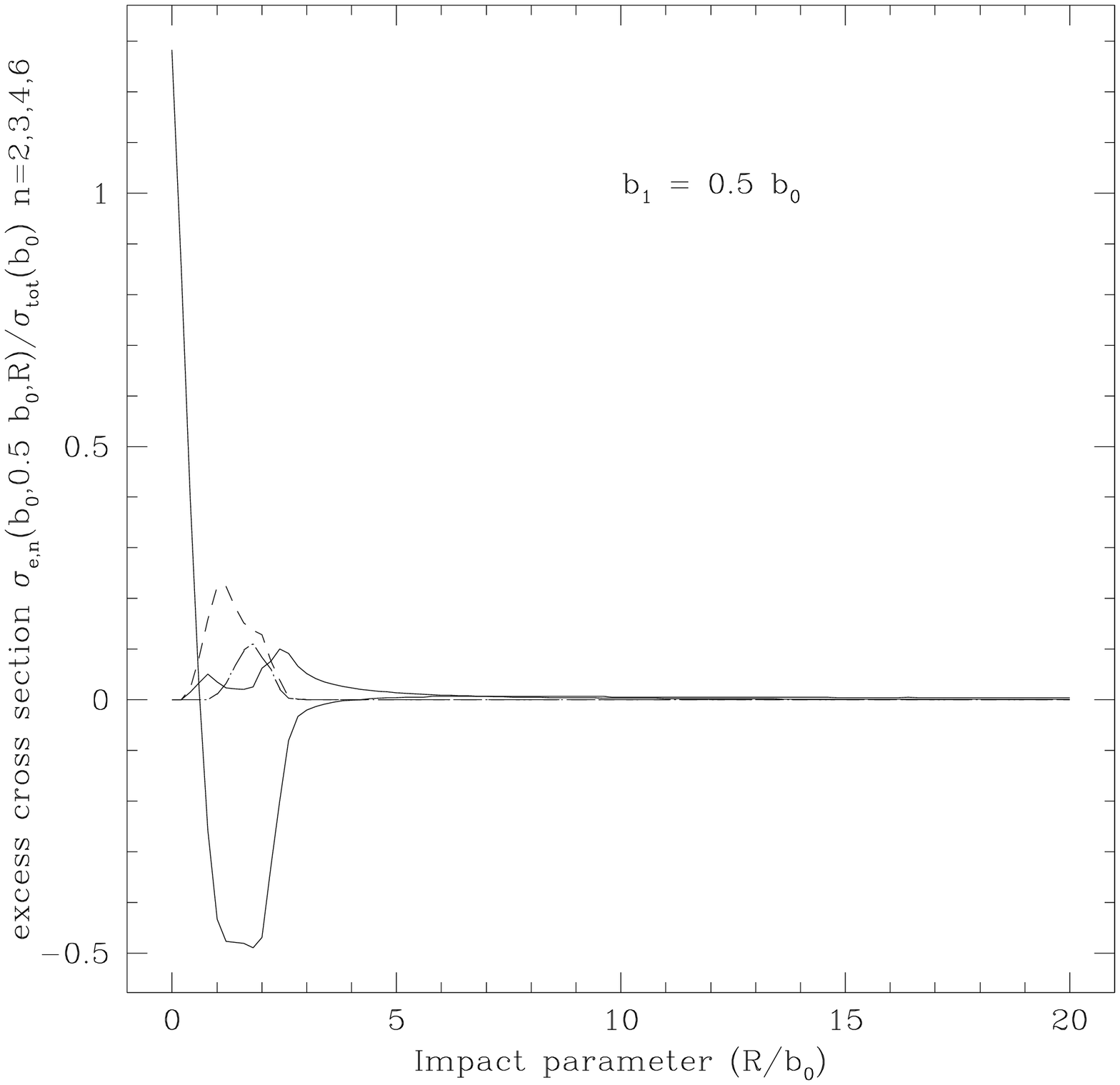}{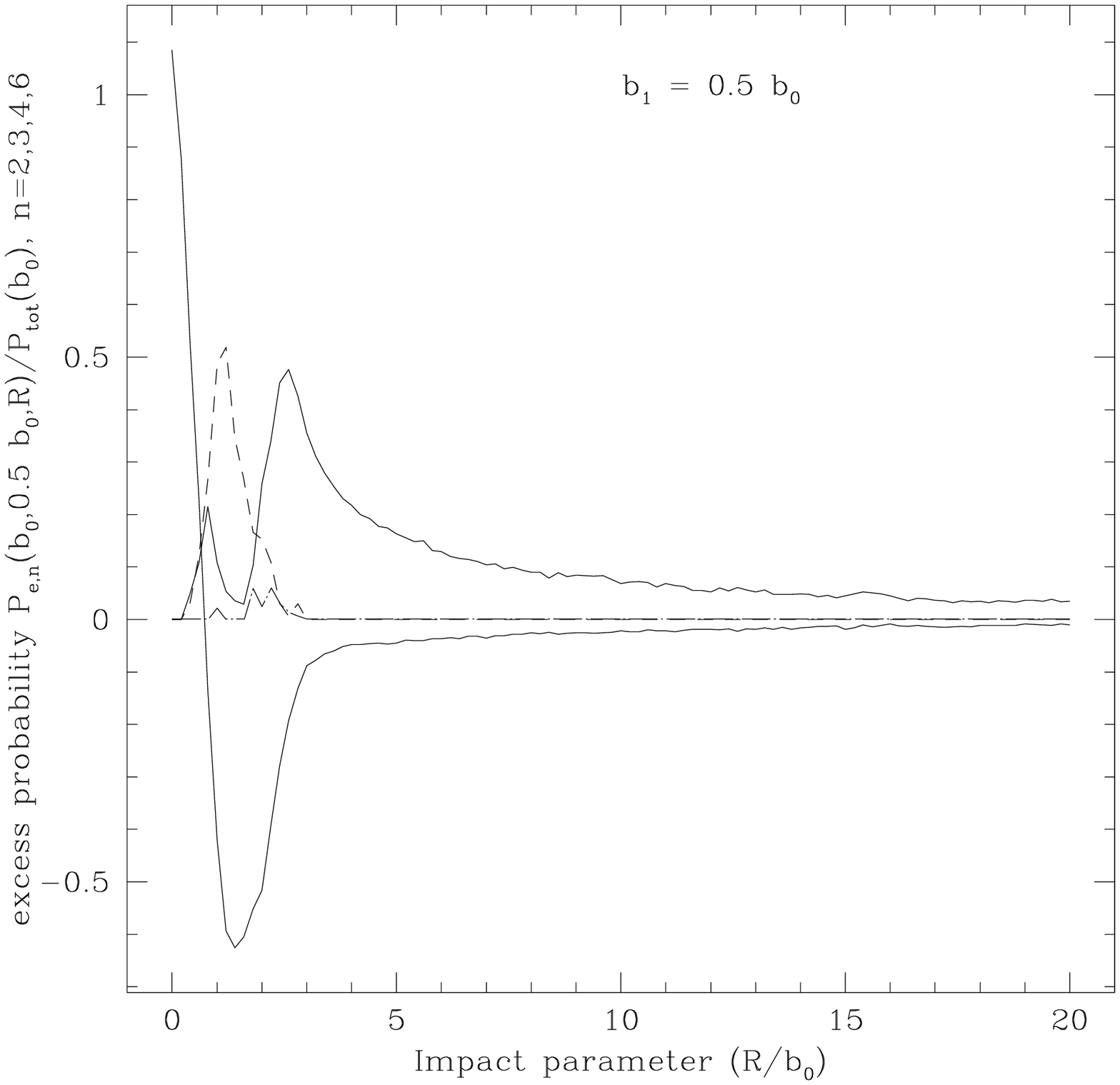}
\caption{
(Left) The excess cross section $\sigma_{e,n}(b_0,b_1,R)$ for finding $n=2$, 
$3$, $4$,
and $5$ images as a function of impact parameter $R/b_0$ for $b_1/b_0=0.5$. 
(Right) The excess lensing probability $P_{e,n}(b_0,b_1,R)$ 
for finding $n=2$, $3$, $4$
and $5$ images as a function of impact parameter $R/b_0$ for $b_1/b_0=0.5$.  
In both, the solid line dipping below zero refers to doubles,
triples are denoted with a dashed line,
quads with a solid line, and fives with a dot-dashed line.
}
\label{fig:excess1}
\end{figure}

We simulate the procedures of the CLASS radio survey.  To be selected for 
further study, the lens has to have at least one image pair with a separation
larger than $0\farcs3$ and a flux ratio smaller than 15:1.  Neither of 
these criterion has an enormous effect on the detectability of our model
lenses, although for low satellite masses 
(i.e. when $b_1 < 0\farcs15=0.22b_0$), 
we detect only the images dominated by the larger deflection scale of the
primary lens.  These selection functions have little effect; we lose
only 1\% of the lenses for large satellites and about 4\% of lenses
for $b_1/b_0=0.5$.  
We assumed that all lenses selected for follow-up studies would undergo
VLBI observations that could detect all images with separations larger
than $0\farcs01$ and flux ratios within 100:1 of the brightest image.
Ignoring the images in the lens cores, we used this criterion to 
set the number of visible images $n$ we used to classify the lenses.

If the lenses are well separated, the images are associated with one
or the other lens and the effects of the companion are
well-approximated by an external shear perturbation of 
amplitude $\gamma=b/2R$.  
Once this shear perturbation becomes smaller than other sources of
shear on the lens (the ellipticity of the lens or the shears from more
distant halos), a model based on two circular lenses becomes unrealistic.
Thus, we truncate our radial integrations once the induced shears
drop below a level of $\gamma=0.025$.  For satellites massive enough
to produce detectable images ($b_1 > 0\farcs15$), we set the cutoff
radius based on the shear induced by the primary lens on the satellite.  
For lower mass satellites, where the CLASS survey could not resolve
multiple images generated by the satellite, we set the cutoff radius
based on the shear produced by the satellite on the primary.  No
matter the exact definitions, our effective cross sections will have
a discontinuity as the mass of the satellite becomes large enough
to produce directly detectable image separations.

A satellite with a mass too small to produce 
detectable image separations is also too small for us to simulate
(due to numerical resolution limitations).
We thus had to estimate how many of the
lenses we found for larger satellite masses would actually be
detectable for smaller satellite masses.  The way we did this
was to assume that all the lens systems associated solely with the satellites
would be undetectable if twice the satellite Einstein radius,
a characteristic lens separation, was
below the CLASS survey resolution.  ``Associated solely with
the satellites'' means that for instance the two lenses are widely
separated and all the images are located around the satellite.
We estimated the number of images which
would be solely associated with the primary and then extrapolated
this number to low satellite masses, where the images associated solely
with the primary are the only images visible.
There were two ways of estimating these ``primary only'' images.
First, we took only sources within 
3.4 $b_0$ of the primary once the lenses were well separated (varying
this cutoff distance from 2.4 $b_0$ to 4 $b_0$ was a small effect).
This distance was chosen because for all satellite masses of interest
the caustics of the two lenses would not overlap at this separation.
Secondly, we took only lenses which had at least one image on the
``far side'' of the primary lens in the primary-secondary configuration.
We fit the resulting excesses in both cases as a function of $b_1$, 
extrapolated the results for the region with $b_1< 0.22 b_0$, and then
used this in our final estimates.  The two approaches lead to differences
of only 2\% in our estimates of the quad lens fraction, with fewer 
quads found when we cut the sample based on the source position
rather than the image position.  For the remainder of the paper, we
used the source plane cuts because they give the more conservative
(i.e. smaller) effect.

\begin{figure}
\plottwo{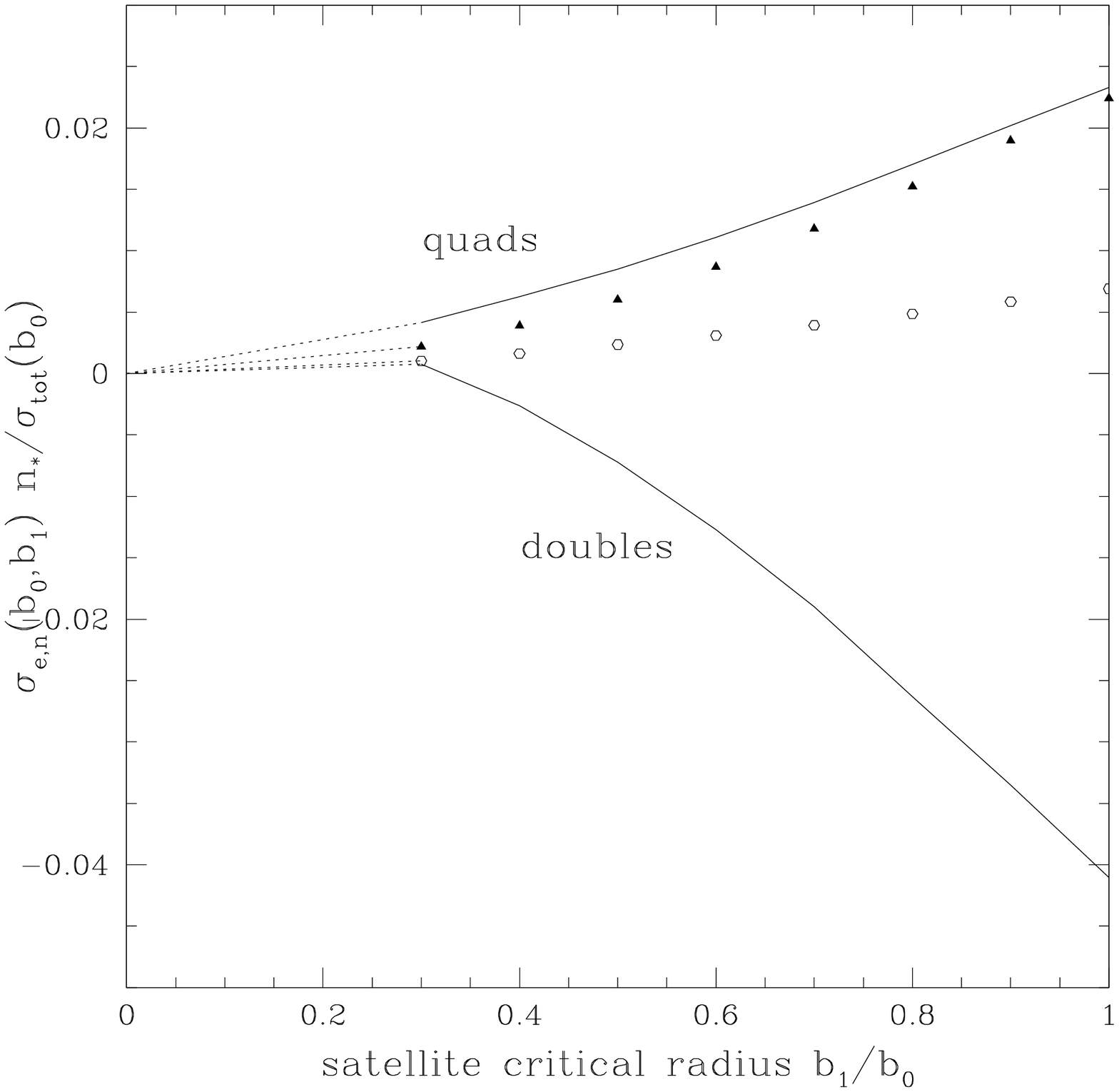}{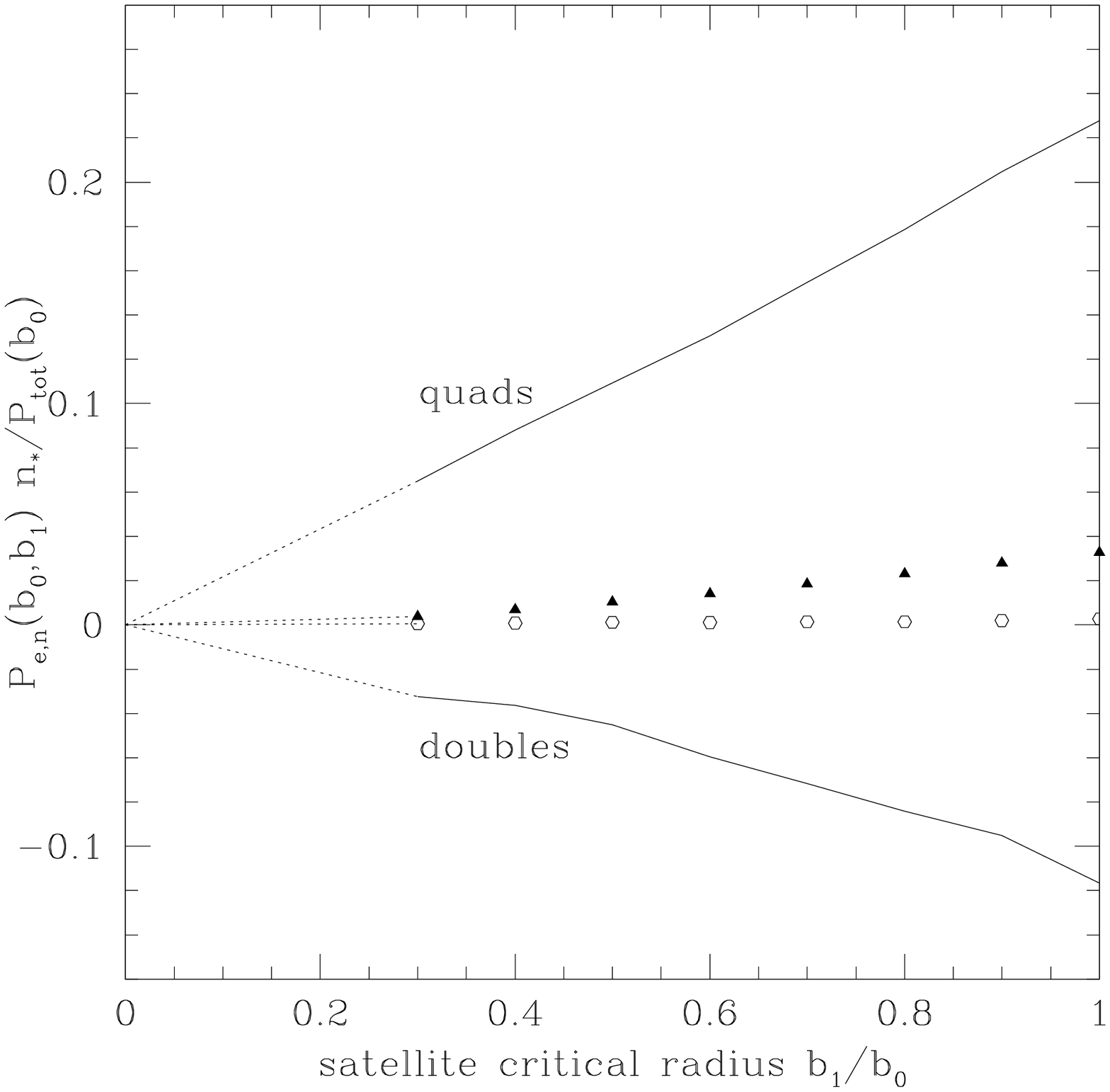}
\caption{
(Left) The excess cross section $\sigma_{e,n}(b_0,b_1) n_*$ for finding 
$n=2$, $3$, $4$
and $5$ images, as in equation \ref{eqn:intdr}.
(Right) The excess lensing probability $P_{e,n}(b_0,b_1) n_*$ 
for finding $n=2$, $3$, $4$
and $5$ images as in equation \ref{eqn:intdr}.
The dashed straight lines connect the smallest
calculated value to the origin, as the excess should
disappear as $b_1 \rightarrow 0$.
In both, the solid line dipping below zero refers to doubles,
triples are denoted with filled triangles,
quads with a solid line, and fives with open circles.
}
\label{fig:excess2}
\end{figure}

\begin{figure}
\leavevmode
\plottwo{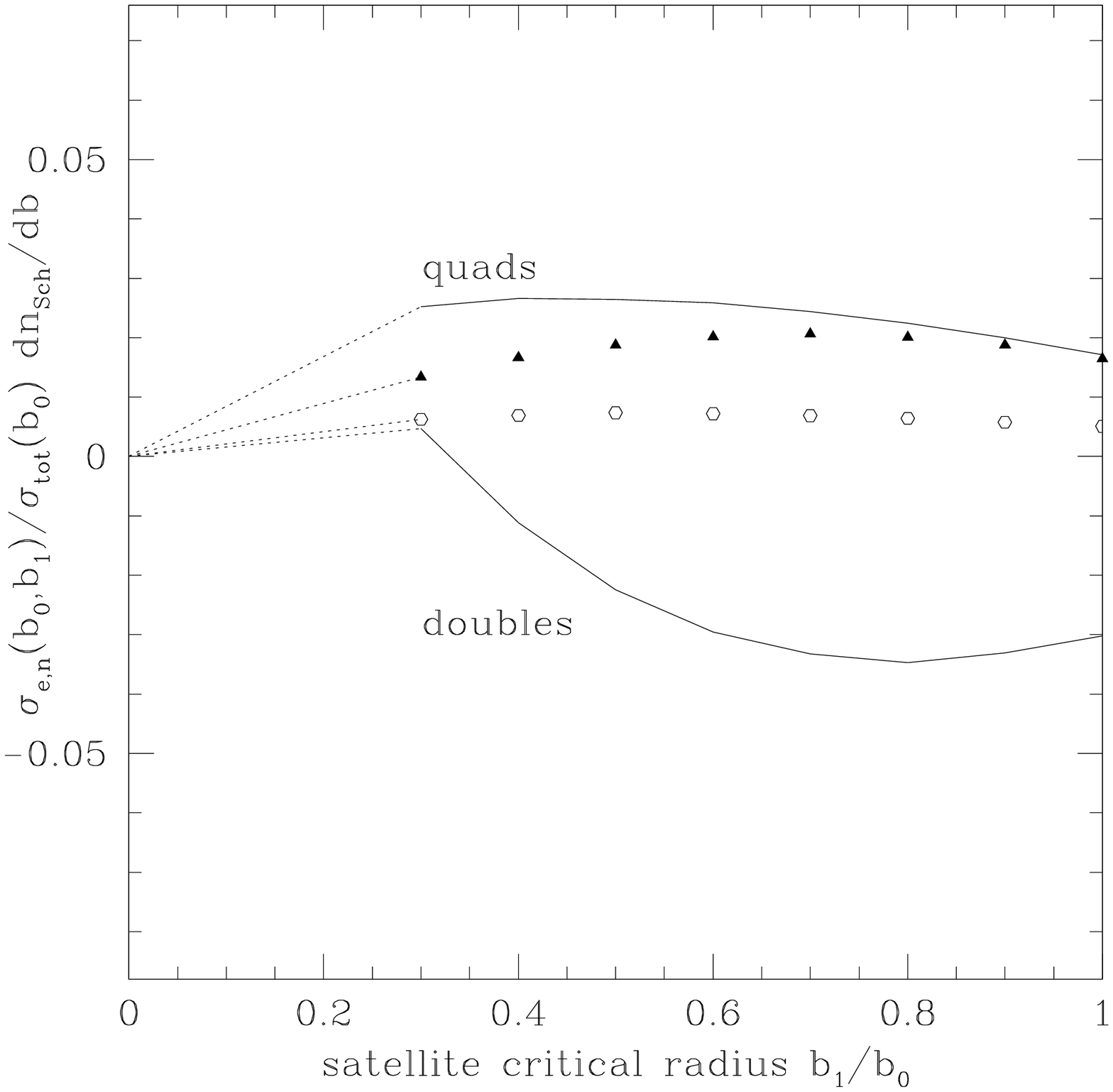}{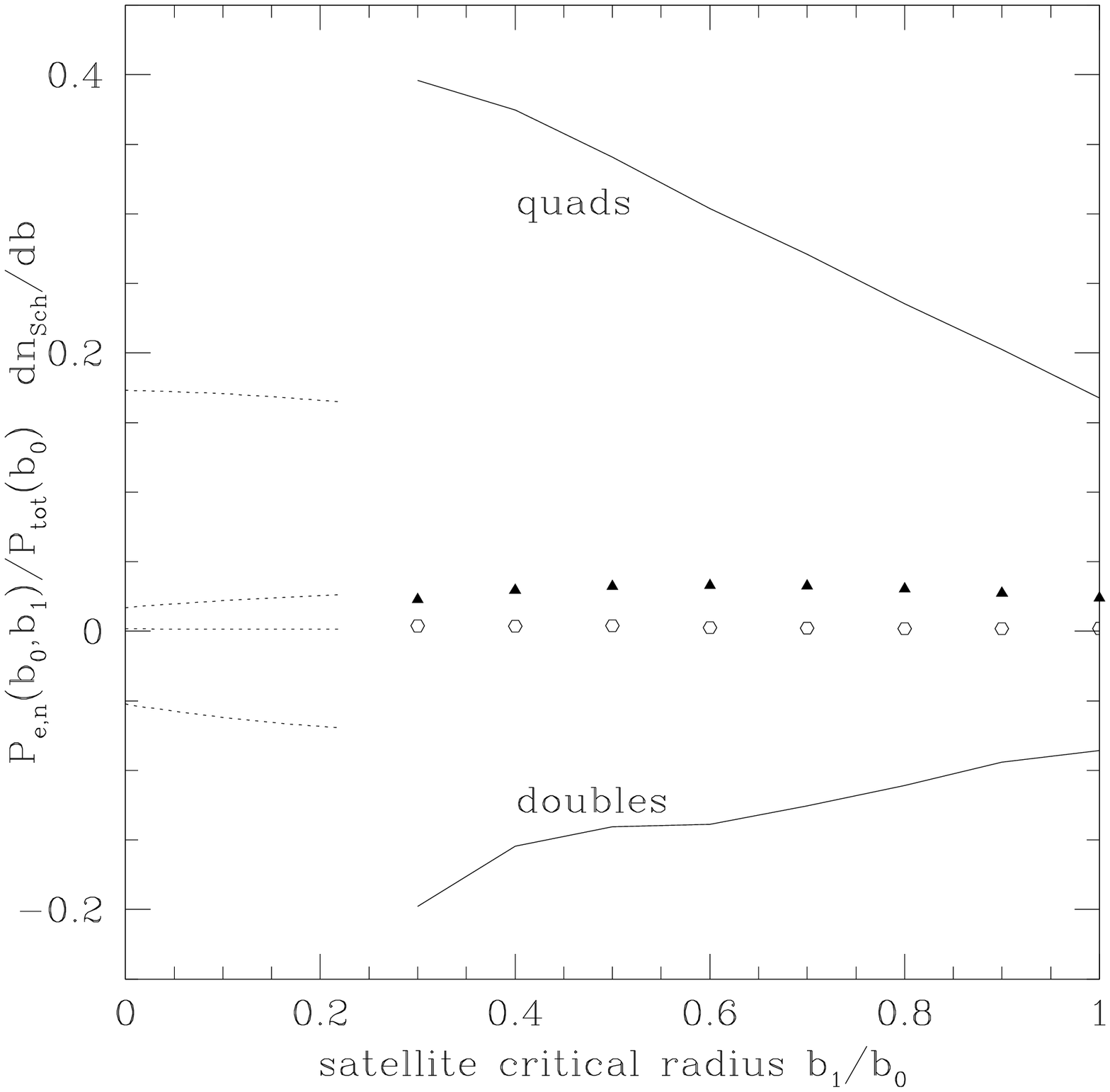}
\caption{
(Left) The excess cross section 
$\sigma_{e,n}(b_0,b_1) (2/b)\exp(-b^2/b_*^2)$ for finding 
$n=2$, $3$, $4$
and $5$ images, left hand side of
equation \ref{eqn:intdr}.  The dotted lines connect the lowest
calculated points from simulations with the origin.
(Right) The excess lensing probability $
P_{e,n}(b_0,b_1) (2/b)\exp(-b^2/b_*^2)$ for finding $n=2$, $3$, $4$ and 
$5$ images, left hand side of equation \ref{eqn:intdr}.
For $b_1\le 0.22b_0$ the angular selection function makes it difficult
to see the lenses produced by the low mass lens at large separations.
This reduces the number of images seen and also requires that
we truncate our integrals at a separation dictated by the shear on the
primary due to the secondary rather than vice versa.   The fit to the
extrapolated truncation of this quantity (counting images associated with
primary only) causes the noticeable break in the curves once the
image separations associated with the satellite are too small to be
seen in the CLASS survey.
In both, the solid line dipping below zero refers to doubles,
triples are denoted with filled triangles,
quads with a solid line, and fives with open circles.
}
\label{fig:excess3}
\end{figure}

\section{Results}

We first consider the scalings expected for our standard model, and then
briefly consider the results for alternative assumptions for the
halo mass function, correlation function, and halo mass profile.
 To provide a sense of the expected numbers, the fraction
of image configurations with $n=4$ visible images produced by an isolated 
lens with an axis ratio of $0.7$, typical of early-type galaxies,
is 3\% based on the cross sections, and 21\% based on the lensing
probabilities.  The fraction is higher when based on the probabilities
because the $n=4$ image systems have higher magnifications and thus
larger magnification bias factors than the $n=2$ image systems.
This estimate roughly corresponds to results
of calculations which average over the ellipticity distributions of 
early-type galaxies (e.g. Kochanek~\cite{Koc96b},
Keeton, Kochanek \& Seljak \cite{kks}, RT, Chae \cite{Cha02}).  
Recall that the observed fraction of lenses with four or more
images in the published JVAS/CLASS sample is
50\% (9 quads and a sextuple in a sample of 20 lenses),
so satellites must roughly double the quad lens fraction if they
are to be a significant part of the solution to this problem.

In Figures~1--3 we illustrate the various excess cross sections and
probabilities.
In Fig.~\ref{fig:excess1} we show the 
excess cross section and 
probability as a function of impact parameter (separation) $R$ for a lens
with $b_1=0.5 b_0$.  As the two lenses are brought together, the
combined lens produces more $n=4$ image lenses and fewer $n=2$ image
lenses as the mutually induced tidal shear converts some of the $n=2$ 
image
cross section into $n=4$ image cross section (and some $n=3$ and $5$ 
cross section).  The depression of the $n=2$ image
cross section is largest at ``resonance'' where the center of each lens
galaxy is projected to the same source point and there is a peak in
the cross section for producing additional images
(see Kochanek \& Apostolakis~\cite{KocApo88}).
As the lenses become still closer, the induced ellipticity diminishes
and there is a net excess of $n=2$ image systems.

Note that the excess probability, the lower panel in
Fig.~\ref{fig:excess1}, has a different asymptotic slope at
large impact parameters from the excess cross section because of the
effects of magnification bias.  The shear produced at lens $1$ by
lens $0$ at separation $R$ is $\gamma=b_0/2R$,
which results in a tidally induced $n=4$ image cross section proportional
to $b_1^2 \gamma^2 \sim b_0^2 b_1^2/4R^2$.  The average magnification
of these images diverges as $M \sim \gamma^{-1}$, so the magnification bias
grows like $B \sim R^\alpha$ where $\alpha \sim 1.1$ is the slope of the
number counts (see Kochanek~\cite{Koc96a}).  
As a result, the excess probability decreases only
as $R^{\alpha-2}$ rather than the $R^{-2}$ scaling of the cross section
at large impact parameter.  This very slow $\sim R^{-1}$ convergence
of the excess $n=4$ image probability can lead to diverging total 
probabilities if the integral over the impact parameters is not 
truncated.  The natural truncation scale is the radius at which the
induced shear is comparable to the typical intrinsic ellipticity of
the lens galaxy or the typical tidal shear, because it is on this
scale that the integrals for (more complete) models 
with additional sources for shear 
would rapidly converge to the results for isolated lenses.  For
a tidal shear $\gamma_t$ the average cross section would be 
$\propto b_1^2 (\gamma_t^2 + b_0^2/(4R^2))$ and the average
magnification would be $\propto (\gamma_t^2 + b_0^2/(4R^2))^{-1/2}$  
leading to rapid convergence of the excess probability to zero once 
$\gamma_t \sim b_0/2R$.

Fig.~\ref{fig:excess2} shows the result of integrating these functions
over the satellite density distribution
\eq
\label{eqn:intdr}
\begin{array}{ll}
   \sigma_{e,n}(b_0,b_1) \frac{dn_{Sch}}{db} & \equiv 2 \pi \int_0^{20 b_0} R_c dR_c { dn \over dA_c db_1} 
                               \sigma_{e,n}(b_0,b_1,R) \\
   P_{e,n}(b_0,b_1) \frac{dn_{Sch}}{db} & \equiv 2 \pi \int_0^{20 b_0} R_c dR_c 
        { dn \over dA_c db_1}
P_{e,n}(b_0,b_1,R)
\end{array}
\en
up to our maximum separation $20 b_0$. 
(Recall $\frac{dn_{Sch}}{db} = \frac{2 n_*}{b} e^{-b^2/b_*^2}$.)
Note the weighting over 
$F(b_1/b_*)$ for the uncorrelated terms so that
the correlated and uncorrelated terms are directly comparable 
(i.e. they will both be
multiplied by $\frac{dn_{Sch}}{db}$ to get the full contribution).  
On a per satellite basis, massive
satellites dominate the excess cross section and probability, just as
they dominate the cross sections of isolated lenses.  If we integrate
over the correlation function with the radial integral truncated at
an inner radius of $R=b_0$ (roughly corresponding to the resonance
region where the production of 5 image systems peaks), then the
induced correlated 5-image cross section 
divided by $\frac{dn_{Sch}}{db}$ is $\propto n_1 r_0^{1.8} b_0^{1.2} b_1^2$.
The uncorrelated term also is proportional to $b_1^2 + O((b_1/b_*)^3)$, 
so the combination roughly matches the shape of the 
curves in Fig.~\ref{fig:excess3}.

These distributions exaggerate the importance of the more massive satellites
because they are not weighted by the relative abundances of high and low
mass systems.  Low mass systems are more abundant, due to the additional
required factor $\frac{dn_{Sch}}{db}$, and so are more
likely to be close to the critical radius of the primary lens and
produce significant perturbations.  
Fig.~\ref{fig:excess3} shows the effect of including this factor.
For example, the product of the uncorrelated cross section ($\propto b_1^2$)
with the Schechter function leads to an overall scaling 
$b_1 \exp(-b_1^2/b_*^2)$ that leads to the rise and then the fall in the
excess cross sections shown in Fig.~\ref{fig:excess3}.

The final integrals were performed using a quadratic fit to the probabilities 
as a function of $b_1$ before including the $dn_{Sch}/db$ weighting (so that we
interpolate a smoother function), with the interpolating function forced
to be zero as $b_1\rightarrow 0$.
When we combine all these effects by integrating over the distribution
of less massive satellites,
\eq 
\begin{array}{l}
\sigma_{e,n}(b_0) \equiv \int_0^{b_0} db_1 \frac{dn_{Sch}}{db}
\sigma_{e,n}(b_0,b_1) \; , \\ 
   P_{e,n}(b_0) \equiv \int_0^{b_0} db_1 \frac{dn_{Sch}}{db}
P_{e,n}(b_0,b_1) \; ,
\end{array}
\en
we find a significant effect.  

There is a net loss in the two-image
cross section of $P_{e,2}(b_0)/P_{tot}(b_0) = -0.11$, a net gain
in the four-image cross section of $P_{e,4}(b_0)/P_{tot}(b_0) =  0.26$,
and a small contribution from other 
multiplicities ($P_{e,3}(b_0)/P_{tot}(b_0) =  0.03$
and  $P_{e,5}(b_0)/P_{tot}(b_0) =  0.003$).   The total cross section
is slightly changed from that of the two potentials in isolation, with a
net increase in the total probability of about 18\%.  This increase
in probability was taken into account in our calculation of $n_*$ as
mentioned earlier.

The effect on the quad fraction is much larger because we combine the effects
of a net suppression in the $n=2$ image cross section with a net gain in the 
$n=4$ image cross section.  If a typical elliptical model predicts that
fraction $f \simeq 21\%$ (for axis ratio 0.7) of lenses will
be four image systems, the addition of satellites should change
the quad fraction to 
\eq
\label{quadfound}
(f P_{tot}+P_{e,4})/(P_{tot}+P_{e,2}+P_{e,3} +P_{e,4}+P_{e,6})\simeq 40\%.
\en
For doubles, we have 
\eq
\label{doubfound}
(f P_{tot}+P_{e,2})/(P_{tot}+P_{e,2}+P_{e,3} +P_{e,4}+P_{e,6})\simeq 57\%.
\en 
The satellites nearly double the expected quad fraction, thereby solving
the quads-to-doubles ratio problem given the statistical uncertainties in
the observed ratio and our estimate of $n_*$.  The model also predicts
a 3\% contribution from $n=3$ and $n=5$ lenses, 
corresponding to an expectation
of one lens with a non-standard multiplicity in the JVAS/CLASS sample.  In
practice, the JVAS/CLASS survey found one non-standard lens, B1359+154, 
a 6 image system formed by a compound lens consisting of 3 lens galaxies.

\begin{figure}
\label{fig:contrib}
\leavevmode
\plottwo{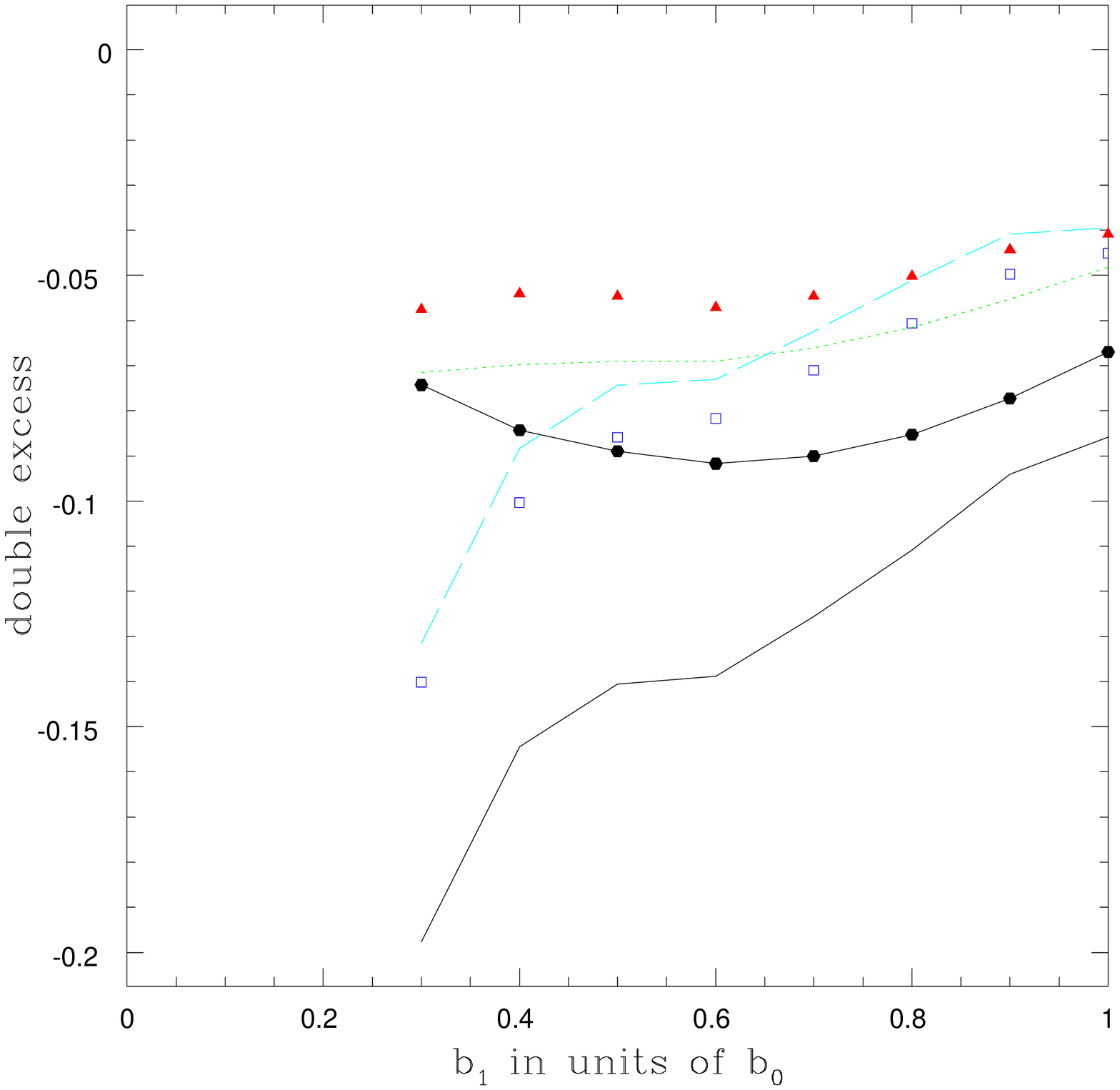}{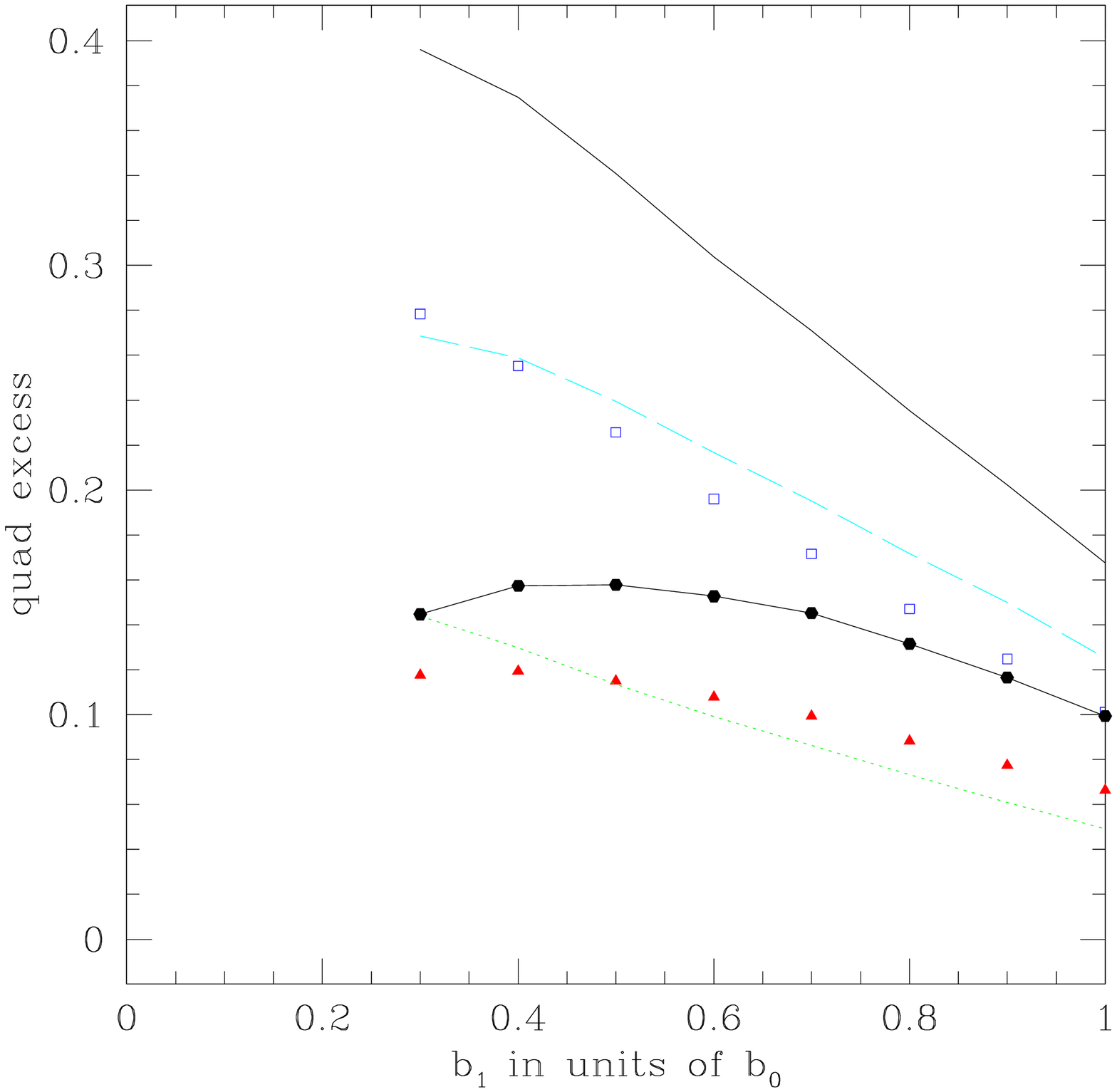}
\caption{
Contributions to doubles (left) and quads (right) from various effects.
The lowest (uppermost) solid line is the full contribution.
The filled triangles are the correlated contribution and the
open squares are the uncorrelated contribution.  The dotted
and dashed lines are the contributions from within a radius
of $5 b_0$ and outside this radius respectively, and the filled hexagons
with the line going through them
show the contribution from the images associated solely
to the primary.  For images associated solely with the primary,
the net total number of systems is lower, so to allow
easier comparison with the
other numbers, this curve has been rescaled to give the same net
number of systems, thus giving the fractional contribution of
these systems. 
}
\end{figure}

If we attempt to break down the excess cross section, we find that
no single source dominates the result.  For massive satellites,
correlated, uncorrelated, nearby ($R<5b_0$) and distant satellites
all make equal contributions. 
This is show in Figures \ref{fig:contrib}.
In addition, lenses due to the primary alone are also shown
(the quad to doubles ratio of the
lenses associated with the primary alone is 31:66).

\begin{figure}
\label{figecomp}
\leavevmode
\plottwo{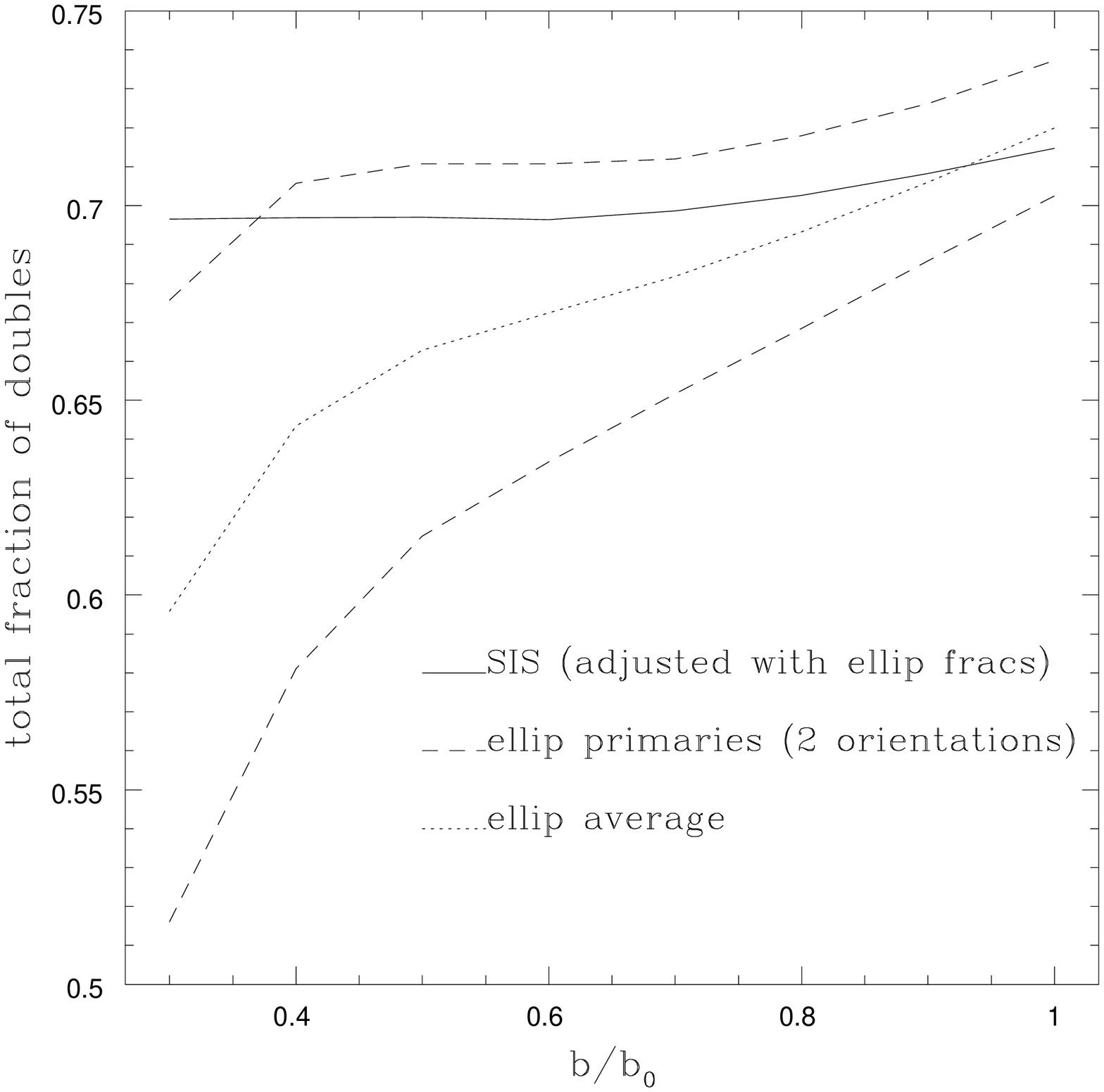}{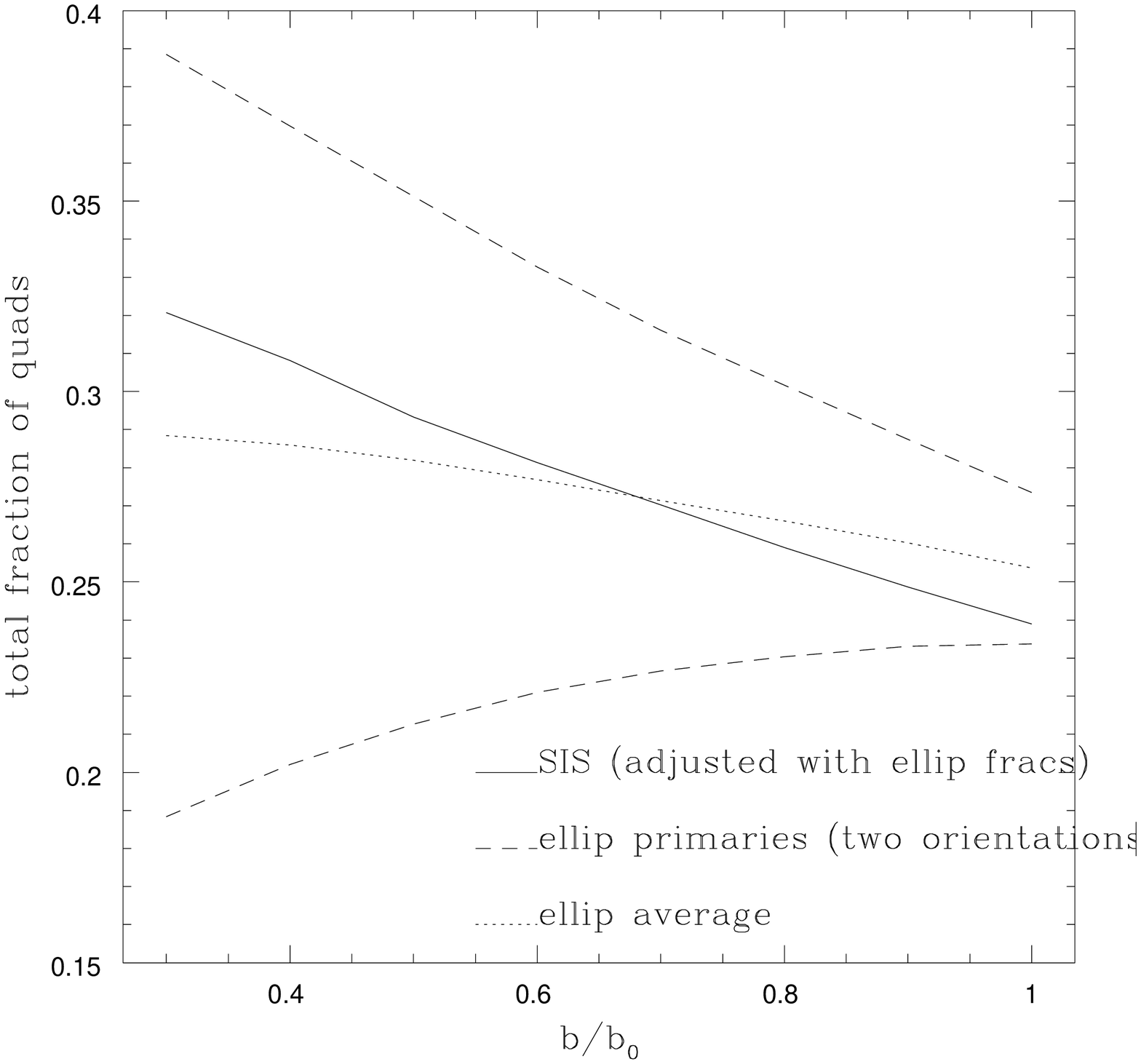}
\caption{Comparison of SIS (with core) primaries and
elliptical primaries.
The fraction of doubles are shown on left, quads on right, plotted
as a function of satellite mass.  
The solid line is the fraction of doubles or quads
in the SIS (with core) case used for the bulk
of the calculations, where the fractions found from
an SIS primary and satellites are added to the fractions 
for an elliptical primary in isolation.  
The dashed lines are for elliptical primary
lenses either aligned parallel or perpendicular to the direction of
the satellite.
The dotted line is the average of the two elliptical orientations.
As the overlapping caustics of the two lenses was expected
to cause the most variation from the ansatz used in the paper, 
separations only up to 5 $b_0$, five times the Einstein radius of 
the primary, were included; at and beyond this point the caustics 
of the two lenses are well separated.  As can be seen in the plots
the approximation used in the bulk of the calculations is quite
close to the average of the two elliptical orientations for the
quad case, for the double case the SIS calculation will actually
tend to give an overestimate.  
}
\end{figure}
In our SIS+SIS model and the computation of the change in the quad
fraction we model the number of two (four) image lenses produced by
a more realistic SIE+SIS model as the sum of the number expected for
an SIE in isolation with the excess cross section.  We compare this
simple model to results for the more realistic SIE+SIS models with
satellites on either the major or minor axis of the lens in 
Figure 5.  The results with the SIE bracket the SIS models 
with an average
quite similar to our simpler spherical model.  If we conducted a 
full suite of models at all inclinations we would simply fill in
the region between the major and minor axis limits.  We can crudely
understand why ignoring the ellipticity of the primary lens has
little effect on the result by considering a lens with two 
independent external shears $\gamma_0$ and $\gamma_1$ representing
the ellipticity of the primary lens and the shear induced by a 
satellite respectively. 
The four-image cross section of the combined lens simply scales as 
$\gamma_0^2 + \gamma_1^2 + 2\gamma_0\gamma_1 \cos 2\Delta\Theta_\gamma$
where $\Delta\theta_\gamma$ is the angle between the two shear
axes.  The angle averaged cross section, $ \sim \gamma_0^2+\gamma_1^2$,
is simply the sum of the four-image cross section of the primary
galaxy in isolation with that induced by the satellite on a
spherical galaxy.  Thus, while a complete calculation with elliptical
primaries averaging over orientation is a logical next step, it should
have little effect on our basic, quantitative conclusion.
We also found that the non-circular models were 4--5 times more efficient
at producing $n=5$ image systems.

True lenses will also have an external shear which can
affect the quad to double ratio.  We expect this shear to be random
with respect to the elliptical axes of the primary and thus
simulated random shears.  Averaging over two shear orientations
with ellipticity led to a net increase in the quad to double ratio,
improving the effects of the satellite galaxies, just as would be expected
from the argument above.
A detailed study of the full parameter space would be interesting
for future work.

\subsection{Variations to Standard Model}

Our standard model was based on treating close satellites of the primary lens
as independent galaxies on larger scales.  Using an empirical 
normalization for their abundance ($n_*$) helps keep the results
from changing drastically under changes of the distribution of 
the satellites in mass or radius.  To see how robust these
results are, we consider here
alternate assumptions about the distribution of the satellites in mass, 
separation and or internal structure.

We first consider changing from a Schechter distribution of satellites, 
motivated by galaxies, to a power law model motivated by CDM halo models.  
Simulations find halo mass functions that are power laws, 
\begin{equation}
 { dn \over dM }= \frac{n_*}{M_*} \left( { M \over M_*} \right)^{-1.8}
\end{equation} 
(e.g. Moore et al. \cite{Moo99}, Klypin et al. \cite{Kly99}).
 with a 
normalization of $n_*=0.01h^3$~Mpc$^{-3}$ at $M_*=10^{12}h^{-1}M_\odot$ 
for a standard ``concordance'' 
model with $\sigma_8=0.9$ (Jenkins et al.~\cite{Jen01}). 
 If we are 
using a model with potentially dark halos, we should avoid normalizations
based on the numbers of visible satellites.
We convert mass to luminosity to critical radii (with distance
dependence) as in section \S 2.

In the CDM picture, the number of low mass halos and galaxies rapidly
diverges towards lower mass.  NFW-like galaxy profiles 
(Navarro, Frenk \& White \cite{NFW},
Moore et al. \cite{Moo99})  are expected when only gravitational forces
are important, but these do not lens efficiently.  Once baryonic cooling
occurs, the halos contract to the very
effectively lensing SIS profiles (e.g. Keeton \cite{Kee98},
Porciani \& Madau \cite{PorMad00}, Kochanek \& White \cite{KocWhi01},
Li \& Ostriker \cite{LiOst02a,LiOst02b}).  In some sense, our Schechter
function model for the satellites already is the correct model for the
halos in which baryons have cooled.  However, as an experiment, we 
also tried using the mass function of halos, cutting off the halos
with such low circular velocities that they may have lost all their
gas when the universe reionized (see Bullock, Kravtsov \& Weinberg 
\cite{BulKraWei00}).  We set this scale to 
$b_1 = 0.05 b_*$ (corresponding to  $v_c \simeq 50$~km/s), and
then set the number of halos based on CDM simulations.
Under these assumptions, we again find a significant suppression of the $n=2$
configurations, with $P_{e,2}/P_{tot}=  -0.46$, a significant gain in the 
$n=4$ configurations, with $P_{e,4}/P_{tot}=1.12$, and a small contribution
from non-standard configurations with $P_{e,3}/P_{tot}=  0.12$ and 
$P_{e,5}/P_{tot}=  0.05$.  For these excess cross sections we would 
expect quads to significantly outnumber doubles with a 73:18 ratio, and 
again to have a small number of triples or quints.

We used an extrapolation of the correlation function, measured on large 
scales, to model the projected density of satellites in the inner 
regions of galaxies.
In practice, large mass satellites which orbit too close to the center of the
primary lens will undergo rapid orbital decay and be destroyed (e.g.
Metcalf \& Madau \cite{MetMad01}).  This can be modeled by setting a scale
$R_{fall}\sim 20$--$30$~kpc for the onset of the rapid decay, and then 
assuming there are no satellites interior to this radius.  This leads to a 
modified correlation function $\xi_2(R)$ which flattens in the center to a 
constant value (roughly $\xi_2(R) \sim 10^3 h^{-1}$ Mpc ) rather
than continuing to rise as a power-law.  However, since we must set the 
normalization $n_*$ to reproduce the observed numbers of satellites 
(finding $n_*=0.027\pm0.009 h^3$~Mpc$^{-3}$ for the $R_{fall}=22$~kpc
case we tried as an experiment), the change in the radial structure has
little effect on the lens statistics.  Assuming the Schechter model for the
satellite mass function, we find the now familiar suppression of the 
doubles, $P_{e,2}/P_{tot}= -0.16 $, and enhancement of the quads,
$P_{e,4}/P_{tot}=  0.41$, leading (with the inclusion of
the triples and fives) to an expectation of a quad:double ratio
of 49:49 assuming lenses with a typical axis ratio
of $f=0.7$, considering lenses associated with the
primary alone gives a corresponding ratio of
35:62.  Using the CDM inspired mass function instead of the 
Schechter model produces again significantly 
more quads, 67\%, compared to 30\% doubles.

As the correlation function at short distances is not well known,
we also experimented with changing the logarithmic slope of
$\xi(r) = (r/r_0)^{-\gamma}$ from $\gamma=1.8$.  For each case
we estimated $n_*$ from the observed numbers of satellites. 
Steeper correlation functions weaken the effect of satellites. 
However, a steeper correlation function
can also {\it increase} the number of quads significantly, if it is
coupled with a restriction that no satellites occur within
a fixed radius (such as 22 kpc).  The cause of these correlations
is easily understood from Fig.~\ref{fig:excess1}.  For a fixed
number of satellites, a steeper correlation function simultaneously
adds weight to the central peak and reduces the weight of the 
resonance region, both of which will enhance the probability of
producing two-image lenses.  A shallower correlation function
does the reverse, thereby enhancing the probability of producing
four-image lenses.  Better statistics for the 
distribution of lens galaxy satellites for different image
morphologies may be able to discriminate between these models. 

For our final experiment we truncated the halos of the two lenses by using
pseudo-Jaffe models (Jaffe \cite{Jaf83}, Keeton \cite{Kee01}) with
$\rho \propto b (r^2 + s^2)^{-2} (r^2 + a^2)^{-2}$ and $a=10 b$ rather
than SIS models for the two lens components.  Normalized by the total
cross sections for the isolated pseudo-Jaffe models, the suppression of
the doubles, $P_{e,2}/P_{tot} = -0.11$, and the enhancement of the 
quads, $P_{e,4}/P_{tot} = 0.17$, is somewhat less than for our standard
SIS+SIS models, raising the quad fraction from the 21\% expected for
$f=0.7$ ellipsoids to 35\% rather than to 40\%.

\begin{figure}
 \plotone{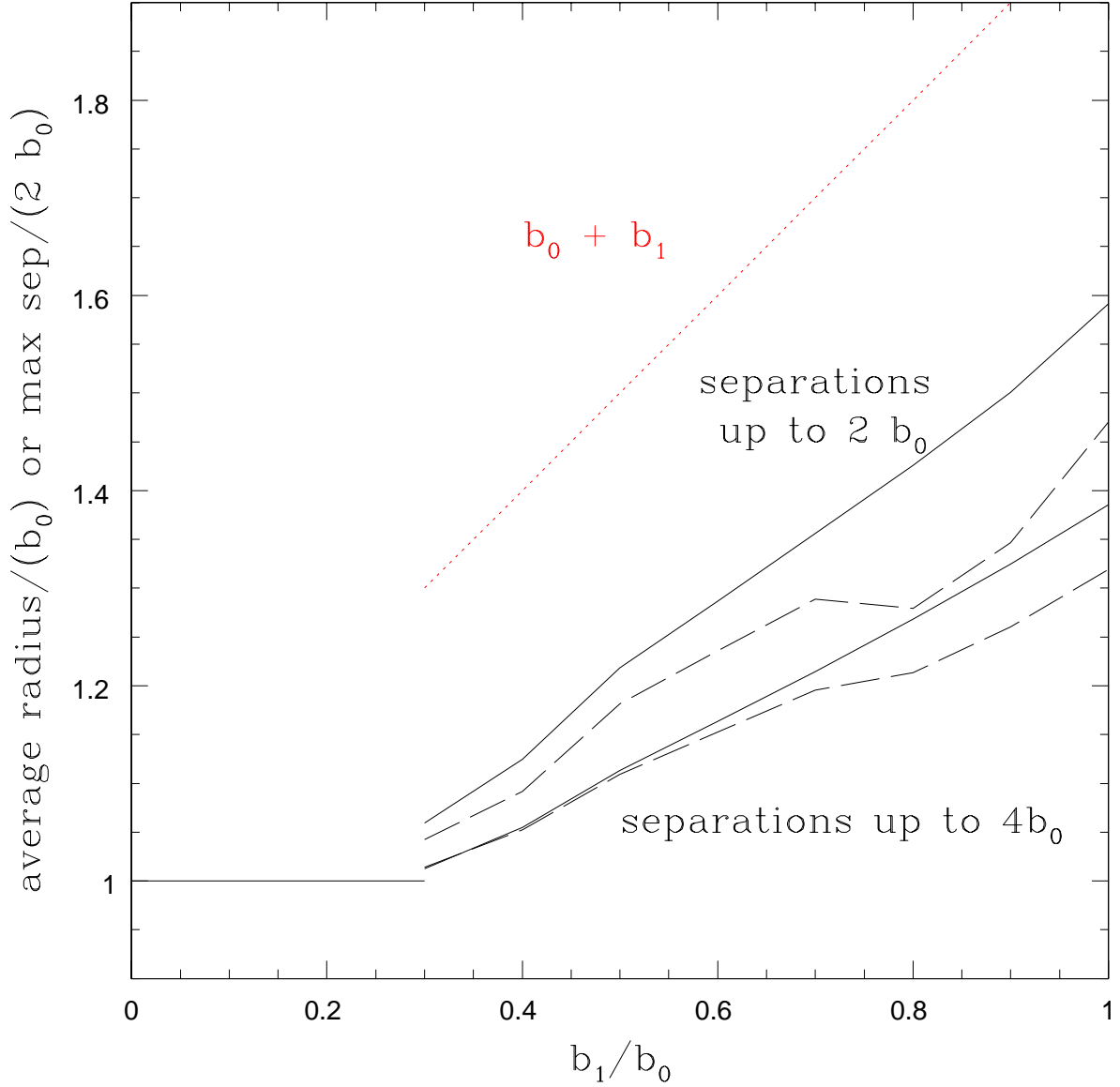}
\caption{Biases in image separation estimates due to satellites.  We
show the average image radius (dashed line) and one-half the maximum
image separation (solid line) in units of the critical radius of
the primary lens $b_0$ as a function of the satellite 
critical radius $b_1$.  We show the averages for regions with
$R \leq 2b_0$ and $4b_0$.The dotted line shows the image
critical radius $b_0+b_1$ which we would find if two SIS
lenses are exactly aligned ($R=0$).
}
\label{fig:sepbias}
\end{figure}

\section{Conclusions}

The high fraction of four-image lenses (50\% of JVAS/CLASS lenses have four
or more images) compared to the expectations for simple ellipsoidal lenses 
in reasonable tidal fields (20\% to 30\%) has been a long standing puzzle
in understanding the statistics of gravitational lenses  (King \& Browne 
\cite{KinBro96}, Kochanek \cite{Koc96b}, Keeton, Kochanek \&
Seljak \cite{kks}, Finch et al. \cite{Fin02}, 
Rusin \& Tegmark \cite{RusTeg01}).  We
find that the problem is largely explained by the changes in the caustic
structures produced by including the effects of nearby satellite galaxies
when we normalize the abundance of these satellites to match the abundance
of satellites in the JVAS/CLASS lens sample.  The satellites 
produce relatively small changes in the total lensing probabilities, but
systematically suppress the probability of obtaining a two-image lens 
in favor of finding a four-image lens, with a small probability of 
finding non-standard 3 or 5 image lenses.  Depending on the parameters
and the assumptions, satellites can roughly double the 21\% quad fraction 
expected for a typical elliptical galaxy (axis ratio $0.7$) up to a 
35--50\% quad fraction that is relatively easy to reconcile with the
data.  Provided we normalize the satellite abundance to the observations, 
this conclusion is little affected by the details of what we assume for the
structure or spatial distribution of the satellites.
A steeply rising mass function will have a notable effect, however,
it is possible that the lowest mass satellites will be affected more
strongly by tidal stripping and be less well approximated by the
Faber-Jackson relation we assumed initially.  It would be interesting
to pursue this in a future work.
The typical satellites responsible for the changes should have critical radii 
(luminosities) roughly 50\% (25\%) that of the primary lens, again 
consistent with the observations where the satellites in the JVAS/CLASS
sample have luminosity ratios of $0.08$, $0.09$, $0.26$, $0.40$, $0.40$, 
$0.45$ relative to the primary lenses.  The surveys should also find that
$\sim$3\% of the lenses should have non-standard multiplicities, which
is borne out by the discovery of two higher multiplicity lenses
(B1359+154, a 6 image lens produced by 3 galaxies in the CLASS survey,
Myers et al. \cite{Mye99}; and PMNJ0134--0931, a 5 image lens produced
by 2 galaxies in the PMN survey, Gregg et al. \cite{Gre02},
Winn et al. \cite{Win02}) in the flat-spectrum radio lens surveys.

The satellite galaxies can also interfere with efforts to estimate the 
cosmological model using lens statistics (e.g. Chae et al. \cite{Chaetal02}, 
Davis, Huterer \& Krauss~\cite{Davis02} and references therein),  as 
it is crucial in such studies to 
match the observed distributions of image separations in order to correctly 
estimate the lens cross sections (see Kochanek \cite{Koc96a}).  Close
satellites modify the image separations, leading to biased estimates of
the average multiple-imaging cross sections that in turn bias estimates
of the cosmological model.  Simply dropping lenses with satellites is 
not a solution because it builds a bias against high mass lenses into
the calculation.  The changes in the cross section produced by the
satellite depend on the ratio $r=b_1/b_0$, while the luminosity of the
satellite depends on $b_1^2=r^2 b_0^2$.  Thus, even though the satellites
create the same fractional perturbation for all lenses, the satellites of
the higher mass lenses are more easily detected simply because they are
more luminous.  For example, suppose every lens of luminosity $L_0$ had a
satellite of luminosity $L_1=L_0/4$, and that we could detect any 
satellite with $L_1 > L_*/4$.  Every lens with $L_0 >L_*$ would be
rejected from the sample because it has a detectable satellite. The
smaller average image separations of the final sample (only 68\% that
of the original sample) would lead to an underestimate of the average
velocity dispersion (by about 20\%) and cross section (by a factor of
two) of the lenses.  {\it But the bias on the cross sections produced
by the existence of satellites has not changed!}All the lenses remaining
in the sample have satellites producing the same fractional perturbations
to the cross sections as they do in the rejected, massive systems.  This
is an extreme example, but the resulting bias in more realistic cases still
represents a serious problem, and may explain the very low velocity 
dispersion scales found by Chae et al. (\cite{Chaetal02}) and 
Davis et al.~(\cite{Davis02}).

Inclusion of the satellites does bias the image separations upwards,
however, as shown in Fig.~\ref{fig:sepbias}.
We computed the average distance of the lensed
images from the primary lens or one-half the maximum image separations,
standard estimators for the average critical radius of a lens, as a
function of the critical radius of the satellite.  We included the
weighting of the radial distribution of the satellites by the correlation
function and computed the averages for the regions with $R \leq 2b_0$
and $4b_0$.  The amplitude of the bias is modest except for systems
with comparable masses.  In particular, note that the average effect
is significantly smaller than simply adding the critical radii of the
two lenses.  If we continue our (extreme) thought experiment of a sample of 
lenses all having satellites with $b_1/b_0=1/2$, we would overestimate
the average image separations of an isolated lens by approximately 20\% 
(using the $R<2b_0$ region, for a more detailed calculation one would 
choose this region self-consistently) if we simply used the observed image
separations. For this case we would get an overestimate of the velocity 
dispersion (by about 10\%) and the cross section (by 40\%).  {\it However, 
the magnitude of the bias from simply including all the systems with 
satellites is only half that from excluding them. } 

In practice, the magnitude of the effect and the ambiguities arising from 
any simple treatment mean that reliable estimates of cosmological 
parameters or galaxy mass scales from lensing statistics cannot be
based on isolated lenses.  The calculations must include the effects
of satellites.  Fortunately, the model for the satellites, particularly
the abundance of satellites, can be calibrated directly from the
observations.  As part of this process, more careful observational
surveys need to be made of lens galaxy environments than the crude
normalization estimates we made in \S2.2. 
 Alternatively the models can be checked against
numerical simulations of lens environments 
(e.g. White et al. \cite{Whiteal01},
Holder \& Schechter \cite{HolSch02}, Chen, Kravtsov \& Keeton
\cite{CheKraKee03}).  Full calculations with elliptical lenses,
large scale tidal shear fields and satellites should be done.  We
also limited our calculation to satellites, but more attention should
be given to the statistical effects of embedding the lenses in more
massive group or cluster halos.  As the lens sample continues to  
grow, so does our ability to understand and quantitatively model
the effects of satellites, so the ambiguities in statistical studies
of lenses introduced by satellites will be resolved.

\section*{Acknowledgments}
We thank N. Dalal, M. Davis, S. Gaudi, C. Keeton, C.P. Ma, B. Metcalf, 
C. McKee, P. Natarayan, 
D. Rusin, S. Stahler, R. Weschler and especially M. White for comments and discussions.  We also thank the anonymous referee for many helpful suggestions.
JDC was funded in part by 
NSF-AST-0074728 and NSF AST-0205935, and thanks the Aspen Center for Physics
and the Kavli Institute for Theoretical Physics (supported in part by
NSF99-07949) for hospitality during the completion of this work.
CSK is supported by the Smithsonian Institution and NASA grant
NAG5-9265.


\begin{thebibliography}{99.}
\bibitem[1996]{barkana} Bar-Kana, R., 1996, ApJ 468, 17
\bibitem[1987]{BlaKoc87} Blandford, R.D. \& Kochanek, C.S., 1987, Ap. J, 321, 
658
\bibitem[1984]{Blu84} Blumenthal, G.R., et al, 1984, Nature, 341, 517
\bibitem[1986]{Blu86}  Blumenthal, G.r., Faber, S.M., Flores, R., 
Primack, J.R.,
1986, ApJ, 301,27
\bibitem[1998]{Bro98} Browne, I.W.A., Wilkinson, P.N., Patnaik, A.R., Wrobel, J.M., 1998, MNRAS 293, 257
\bibitem[2002]{Bro02} Browne, I.W.A., et al., 2003, MNRAS 341, 13 
\bibitem[2000]{BulKraWei00} Bullock, J.S., Kravtsov, A.V., Weinberg, D., 2000,
ApJ 539, 517
\bibitem[2003]{Cha02} Chae, K.-H., 2003, MNRAS 346, 746
\bibitem[2002]{Chaetal02} Chae, K.-H., 2002, Phys.Rev.Lett. 89, 151301
\bibitem[2003]{CheKraKee03} Chen, J., Kravtsov, A.V., Keeton, C.R.,
2003, ApJ 592,24
\bibitem[2002]{Chi02} Chiba, M., 2002, ApJ 565, 17
\bibitem[2002]{DalKoc01} Dalal, N., \& Kochanek, C.S., 2002, ApJ 572, 25 
\bibitem[2003]{Davis02} Davis, A.N., Huterer, D., \& Krauss, L.M.,
2003, MNRAS 344, 1029
\bibitem[2003]{Evans03} Evans, N.W., \& Witt, H.J., 2003, MNRAS 345, 1351
\bibitem[1976]{FabJac76} Faber, S.M., \& Jackson, R.E., 1976, ApJ 204,668
\bibitem[2002]{Fin02} Finch, T.K., et al, 2002, ApJ 577, 51 
\bibitem[2002]{Gre02} Gregg, M., et al, 2002, ApJ 564, 133
\bibitem[2003]{HolSch02} Holder, G., Schechter, P., 2003, ApJ 589, 688 
\bibitem[1983]{Jaf83} Jaffe, W., 1983, MNRAS, 202, 995
\bibitem[2001]{Jen01} Jenkins, A., et al, 2001, MNRAS, 321, 372
\bibitem[1998]{Kee98} Keeton, C.R., thesis.
\bibitem[1997]{kks}Keeton, C.R., Kochanek, C.S. \& Seljak, U., 1997,ApJ,483,604
\bibitem[2001]{Kee01} Keeton, C.R., 2001, Catalogue of Lens Models,
astro-ph/0102341
\bibitem[2003]{Kee03} Keeton, C.R., 2003,ApJ 584, 664
\bibitem[2003]{KeeGauPet02} Keeton, C.R., Gaudi, B.S., Petters, A.O.,
2003, ApJ 598, 138
\bibitem[1996]{KinBro96} King, L.J., \& Browne, I.W.A., 1996, MNRAS, 282,67
\bibitem[1999]{Kin99} King, L.J., Browne, I.W.A., 
Marlow, D.R., Patnaik, A.R., Wilkinson, P.N., 1999,
MNRAS 307, 225
\bibitem[1999]{Kly99} Klypin, A., et al,  1999, ApJ, 522, 8
\bibitem[1988]{KocApo88}
Kochanek, C.S., \& Apostolakis, J., 1988,
MNRAS 235, 1073
\bibitem[1996a]{Koc96a} Kochanek, C.S., 1996, ApJ, 466, 638 
\bibitem[1996b]{Koc96b} Kochanek, C.S., 1996, ApJ, 473, 595
\bibitem[2001]{Kocetal01} Kochanek, C.S., et al., 2001, ApJ 560, 566
\bibitem[2001]{KocWhi01} Kochanek, C.S., White, M., 2001, ApJ 559, 531
\bibitem[2002]{KocDal02} Kochanek, C.S., Dalal, N., 2002,
ApJ 572,25
\bibitem[2003]{KocDal03} Kochanek, C.S., Dalal, N., 2003,
preprint [astro-ph/0302036]
\bibitem[1987]{kovner} Kovner, I. 1987, ApJ 316, 52
\bibitem[2000]{LiOst02a}  Li, L.-X., Ostriker, J.P., 2002,  ApJ 566, 652
\bibitem[2003]{LiOst02b} Li, L.-X., Ostriker, J.P., 2002, ApJ 595, 603
\bibitem[2000]{Lov00} Loveday, J., 2000, MNRAS 512, 557
\bibitem[1998]{MaoSch98} Mao, S., Schneider, P., 1998, MNRAS 295, 587
\bibitem[2001]{MetMad01} Metcalf, R. B., \& Madau, P.,
2001, ApJ 563, 9 
\bibitem[1998]{MoMaoWhi98} Mo, H.J., Mao, S., White, S.D.M., 1998,
MNRAS, 295,319
\bibitem[2001]{MolBla01} Moller, 0. \& Blain, A.W., 2001,
MNRAS 327, 339
\bibitem[1999]{Moo99} Moore, B., et al., 1999, MNRAS, 310, 1147
\bibitem[1995]{Mye95} Myers, S.T., et al., 1995, ApJ 447L, 5
\bibitem[1999]{Mye99} Myers, S.T., et al., 1999, AJ 117, 2565
\bibitem[2003]{Mye02} Myers, S.T., et al., 2003, MNRAS  341, 1
\bibitem[1996]{NFW} Navarro, J.F., Frenk, C.S., White, S.D.M., 1996, ApJ 462,
563
\bibitem[1992]{Pat92} Patnaik, A.R., Browne, I.W.A., Wilkinson, P.N., \& Wrobel, J.M., 1992, MNRAS 254, 655
\bibitem[2001]{Pea01} Peacock, J.A., {\bf Cosmological Physics},
(Cambridge University Press, 2001, Cambridge) 
\bibitem[2000]{PorMad00} Porciani C., Madau, P. 2000, ApJ, 532, 679 
\bibitem[1977]{ReeOst77} Rees, M.J., Ostriker, J.P., 1977, MNRAS, 179, 541
\bibitem[2001]{RusTeg01} Rusin, D., Tegmark, M., 2000, 
ApJ 553, 709
\bibitem[2002]{Rus02} Rusin, D., et al., 2002, MNRAS 330, 205
\bibitem[2003]{Rus03} Rusin, D., et al., ApJ 587, 143
\bibitem[1976]{Sch76} Schechter, P., 1976, ApJ 203, 297
\bibitem[1993]{SchMoo93} Schechter, P., Moore, C.B., 1993, AJ 105, 1
\bibitem[2002]{SchWam02} Schechter, P., Wambsganss, J., 2002,
 ApJ 580, 685
\bibitem[1992]{SeiSch92} Seitz, S., Schneider, P., 1992, A \& A, 265, 1
\bibitem[1977]{Sil77} Silk, J., 1977, ApJ, 211, 638
\bibitem[1984]{TurOstGot84} Turner, E.L., Ostriker, J.P., Gott, J.R., 1984, 
ApJ, 284,1
\bibitem[1978]{WhiRee78} White, S.D.M., Rees, M., 1978, MNRAS, 183,341
\bibitem[2001]{Whiteal01} White, M., Hernquist, L, Springel, V, 
astro-ph/0107023
\bibitem[1998]{Wil98} Wilkinson, P.N., Browne, I.W.A., Patnaik, A.R., 
Wrobel, J.M., Sorathia, B., 1998, MNRAS 300, 790
\bibitem[2002]{Win02} Winn, J., et al, 2002, ApJ 564, 143
\bibitem[1998]{xanetal98} Xanthopoulos et al, 1998, MNRAS 300, 649
\end{thebibliography}
\end{document}